%% file: aanda.tex
\documentclass{aa}

\usepackage{lastpage}
\usepackage[colorlinks=true,citecolor=blue,linkcolor=blue,urlcolor=blue, backref=false,pdfborder={0 0 0}]{hyperref}
\usepackage{url}
\usepackage{mathrsfs}
\usepackage{natbib}
\usepackage{graphicx}
\usepackage{txfonts}
\usepackage{multirow}
\usepackage{color}
\usepackage[table,xcdraw]{xcolor}
\usepackage{orcidlink}
\usepackage{tikz}
\usepackage{txfonts}
\usepackage{subcaption}
\usepackage{orcidlink}
\usepackage{academicons}
\usepackage{textcomp}
\usepackage{hhline}
\usepackage{adjustbox}

\input{newcomands.tex}

\usepackage[normalem]{ulem}
\bibpunct{(}{)}{;}{a}{}{,} 

\begin{document}
	\title{Fast simulation mapping: from standard to modified gravity cosmologies using the bias assignment method}
    \titlerunning{J.E. Garc\'ia-Farieta et al.}
    \author{J.E. Garc\'ia-Farieta,
    \inst{1,2}\orcidlink{0000-0001-6667-5471}\thanks{jorge.farieta@iac.es},
          Andr\'es Balaguera-Antol\'{\i}nez\inst{1,2} \orcidlink{0000-0001-5028-3035} and
          Francisco-Shu Kitaura\inst{1,2} \orcidlink{0000-0002-9994-759X} 
          }
   \institute{Instituto de Astrof\'{\i}sica de Canarias, s/n, E-38205, La Laguna, Tenerife, Spain \and Departamento de Astrof\'{\i}sica, Universidad de La Laguna, E-38206, La Laguna, Tenerife, Spain}
   \authorrunning{Author}
   \date{Received XX XX, 2024; accepted XX XX, 2024}
  \abstract
   {We assess the effectiveness of a non-parametric bias model in generating mock halo catalogues for modified gravity (MG) cosmologies, relying on the distribution of dark matter from either MG or \lcdm.}
   {We aim to generate halo catalogues that effectively capture the distinct impact of MG, ensuring high accuracy in both two- and three-point statistics for comprehensive analysis of large-scale structures. As part of this study  we aim at investigating the inclusion of MG into non-local bias to directly map the tracers onto \lcdm\  fields, which would save many computational costs.}
   {We employ the bias assignment method (\texttt{BAM}) to model halo distribution statistics by leveraging seven high-resolution \texttt{COLA} simulations of MG cosmologies. Taking into account cosmic-web dependencies when learning the bias relations, we design two experiments to map the MG effects: one utilising the consistent MG density fields and the other employing the benchmark \lcdm\ density field.}
   {BAM generates MG halo catalogues from both calibrations experiments excelling in summary statistics, achieving a $\sim 1\%$ accuracy in the power spectrum across a wide range of $k$-modes, with only minimal differences well below 10\% at modes subject to cosmic variance, particularly below $k<0.07$ \hMpc. The reduced bispectrum remains consistent with the reference catalogues within 10\% for the studied configuration. Our results demonstrate that a non-linear and non-local bias description can model the effects of MG starting from a \lcdm\ dark matter field.}
   {}
   \keywords{cosmology: -- theory - large-scale structure of Universe}
   \maketitle

\section{Introduction}\label{sec:introduction}
Modified gravity (MG) theories serve as the foremost and straightforward alternative framework to the standard cosmological model \lcdm, providing a means to address one of its key conjectures - the late-time accelerated expansion of the Universe \citep{Sotiriou_review_fR_2010, Capozziello2011PhR, Joyce_ReviewMG_2016ARNPS}. To accurately explore both small and large scales of the matter distribution, reliable $N$-body simulations are indispensable, which, due to the incorporation of additional degrees of freedom to account for unique effects, pose computational challenges compared to those employed for the \lcdm\ model \citep{Winther2015MNRAS}. Indeed, a comprehensive testing of cosmological probes to assess the potential deviations of gravity from the predicted by General Relativity (GR) demands the construction of accurate MG catalogues. This is particularly essential in meeting the precision and accuracy requirements expected by current galaxy surveys, including the Dark Energy Spectroscopic Instrument (DESI) \citep{DESI2016arXiv161100036D} survey and EUCLID mission \citep{Laureijs2011arXiv1110}.
A variety of methods have been proposed in literature to expedite the prediction of key observables in MG models and circumvent the high computing expense of full $N$-body simulations. Notable approaches involve the parametrisation of fitting functions tailored to accurately reproduce the matter power spectrum \citep{Winther_emul_2019}, the construction of emulators for efficient approximations \citep{Ramachandra_emul_2021, Ruan_emul_2023, Brando_emul_2022, Arnold_FORGE_emulator_2022MNRAS}, as well as approaches based on the cosmology scaling technique introduced by \citet{Angulo_scaling2010}. These techniques have demonstrated the feasibility of casting some of the features of standard cosmology into their counterparts in MG models without the need of comprehensive $N$-body simulations for the latter \citep[see e.g.][]{Mead_MGscaling2015}. While emulators and fitting functions excel in accurately capturing individual observables of the MG catalogues (e.g., mass function, power spectrum, mass-density relations, etc.), these techniques lack the ability to provide an overall view of the summary statistics of the tracer distribution. On a different front, the scaling technique has the ability to generate mock data that incorporate realistic non-linear effects in cosmological model with known background parameters. Specifically, when applied to MG simulations, this technique has shown to be efficient in reproducing MG catalogues with variations of up to $\sim 3\%$ in the matter power spectrum and up to $\sim 5\%$ in the halo mass function (HMF) for scales up to $k=0.1$\,\hMpc\ in Fourier space \citep[][]{Mead_MGscaling2015}. Although this technique provides valuable insights into simulating realistic nonlinear effects in alternative cosmologies, it faces limitations in accurately capturing the screening mechanisms and is unaware to account for environment dependencies, which is critical for MG models. In fact, the overall accuracy of the method will decrease the farther we scale for cosmologies that are away in cosmological parameter space from the original cosmology \citep[see e.g.][]{Contreras2020MNRAS}.

A number of alternative approaches have become increasingly important, since detailed $N$-body simulations are computationally costly to obtain a complete description of the density field in MG scenarios. In light of this, we propose a novel strategy that bypasses the computational demands of direct simulations by relying on the mock construction technique to make use of a smooth large-scale dark matter field obtained from given initial conditions, and populate it with halos (or galaxies) following a bias prescription. There are several methods suggested in the literature to speed-up the construction of mock catalogues while enhancing precision in their clustering statistics, including PEAK PATCH \citep{1996ApJS..103....1B}, \texttt{PINOCCHIO} \citep{2002MNRAS.331..587M}, \texttt{PTHALOS} \citep{2002MNRAS.329..629S}, \texttt{ICE-COLA} \citep{2013JCAP...06..036T,2016MNRAS.459.2327I}, \texttt{PATCHY} \citep{Kitaura_2014MNRAS}, \texttt{QPM} \citep{2014MNRAS.437.2594W}, \texttt{EZmocks} \citep{2015MNRAS.446.2621C}, \texttt{HALOGEN} \citep{2015MNRAS.450.1856A}, among others \citep[see e.g.][]{2003ApJ...593....1B,Angulo:2013gya,2013MNRAS.428.1036M,2015MNRAS.447..437M,2015MNRAS.447..646C,2016MNRAS.459.2118K,2016MNRAS.463.2273F}. Recently, two distinct approaches have been introduced and successfully validated: the non-parametric Bias Assignment Method \citep[referred to as \texttt{BAM} hereafter][]{2019MNRAS.483L..58B,2020MNRAS.493..586P,2023A&A...673A.130B}, and the parametric WebON method \citep[further details provided in][]{2024A&A...683A.215K,2024arXiv240319337C, WebON2024}. Throughout this paper, we mainly focus on the non-parametric method to perform the analysis of the MG density fields.

The \texttt{BAM} approach has demonstrated remarkable precision up to $\sim1\%$ in the power-spectrum at small scales ($k\sim 1$\,\hMpc and $\sim3-6\%$ for typical configurations of the bispectrum when non-local information is included in the mock generation. A comparison of parametric (including  second-order non-local bias) and non-parametric bias mapping techniques using low mass halos from \lcdm\ simulations (on the order of $\sim10^{8}$\,\Msunh), has been conducted by \citet{2020MNRAS.493..586P}. The key finding drawn from that study suggests that non-parametric approaches such as \texttt{BAM} have the ability to replicate the three-point statistics of a halo distribution at low mass scales, where nonlinear clustering and non-local dependencies are likely to dominate. Recently, parametric bias models \citep[][]{2024arXiv240319337C} have achieved a particularly high accuracy by including third-order non-local bias .

The goal of this paper is to assess whether a bias mapping method such as \texttt{BAM}, is capable to reproduce the main features of halos from MG cosmologies given a smooth \lcdm\ dark matter field. We present an alternative approach to address the problem of generating fast and accurate MG catalogues while dealing with the limitations pointed out by previous techniques. This methodology leverages the stochastic and scale-dependent bias description \citep{ Kitaura_2014MNRAS,Kitaura_2015MNRAS, Kitaura_2016MNRAS,Kitaura_2021MNRAS, Balaguera_2020MNRAS,2022MNRAS.512.2245K,2023A&A...673A.130B} to effectively map modified gravity models based on a benchmark high resolution \lcdm\ simulation. The main assumption behind this approach is that the gravitational potential of the MG models can be expressed in terms of the conventional gravitational potential of GR under the correct bias transformation, and therefore, it can produce the adequate cosmic tracer distribution for alternative cosmologies. These two features together result in the following noteworthy characteristics: firstly, we extend the mapping technique to encompass all scales while ensuring compatibility with summary statistics. This broadens the applicability of our method and enhances its accuracy across a wide range of scales. Secondly, we make use of a non-local bias description of the density fields, which captures additional intricacies effects and provides a more comprehensive representation of the underlying density fields while extracting nonlinear and non-local information from a reference simulation.

To reach this goal we employ a set of high-resolution simulations of MG generated with the \texttt{COLA} (COmoving Lagrangian Acceleration) method. The \texttt{COLA} algorithm has shown to perform well enough to produce mock catalogues for BAO analysis \citep{Ferrero_icecola_mocks_2021}. Recently the \texttt{COLA} method has been also evaluated to perform cosmic shear analysis of simulated data of LSST-Y1, serving as source for emulators which has exhibited remarkable fidelity, meeting stringent goodness-of-fit and parameter bias criteria across the prior, offering a promising avenue for extended cosmologies \citep[see][for further details]{2024arXiv240412344G}. This is of particular importance as MG models are prone to show distinctive patterns in the clustering of biased tracers that deviates from \lcdm. These deviations emerge primarily on large scales, where the modifications to gravity vary the growth of density perturbations and the overall matter distribution. 

The road-map of this study is as follows. We run a set of simulations of $f(R)$ models that follow the Hu-Sawicki (HS) parametrisation \citep{HuSawicki_2007}, which is one of the most studied MG models nowadays. The growth of density perturbations in $f(R)$ gravity models is discussed in detail in \S\ref{sec:mgmodel}. The simulations encompass six distinct scenarios, each characterised by a different level of deviation with respect to \lcdm. They were specially designed to include configurations inside and outside of the confidence regions constrained by observations. Therefore, we consider cosmologies that mimic the clustering of \lcdm\ and remain consistent with it, as well as cosmologies that deviate most from it and are already ruled out by observations. The latter case is of interest because it allows us to evaluate the performance of \texttt{BAM} in capturing the MG signatures when considering models that significantly diverge from \lcdm\ predictions. The dark matter (DM) field of the MG simulations, where each particle has a mass of $M_p\approx10^{10}$\,\Msunh, serves as training data for \texttt{BAM}. Similarly, the reference catalogues correspond to the distribution of massive halos, each with at least 83 DM particles, obtained from a Friends-of-Friends algorithm. The MG simulations and training data set are described in detail in
 \S\ref{sec:sims}. Since our aim is to map \lcdm\ into a MG model, we explore two analyses: one using the underlying \lcdm\ DM field to perform the calibration of bias and kernel with \texttt{BAM} to reproduce the reference MG halo number counts; and second, a calibration carried out with the MG DM field of each model to consistently reproduce their reference halo number counts. The methodology, including a brief description of the core of the \texttt{BAM} algorithm, is thoroughly presented in \S\ref{sec:BAMcalibrations}. The results and analysis of the aforementioned scenarios are presented within the same section. Finally, we end-up with a summary and discussion in \S\ref{sec:summaryconclusions}.

\section{Dynamics in $f(R)$ gravity models}\label{sec:mgmodel}
Scalar-tensor theories are among the potential revisions to Einstein's theory of gravity that have received the most attention \citep[for an updated review, see \eg][]{2019RPPh...82h6901K}. Within the plethora of such theories, the $f(R)$ gravity models constitutes one of the most studied limits that adds an extra degree of freedom to GR, where the gravitational action contains, apart from the metric, a scalar field which describes part of the gravitational field. In this sub-set of models, the Einstein-Hilbert action is modified by replacing the Ricci scalar $R$ by a function of other curvature invariants, so that $R \mapsto R+f(R)$. Among the various functional forms of $f(R)$ that have been proposed in the literature \citep[for a review see e.g.][]{Sotiriou_review_fR_2010,DeFelice_review_fR_2010}, we consider the HS template \citep{HuSawicki_2007}, which suggest a plausible function able to satisfy the solar system constraints as well as encode enough freedom to agree cosmic acceleration and structure formation on large scales. The explicit form of HS $f(R)$ function is 
\begin{equation}
f(R)=-m^2\frac{c_1\left(\frac{R}{m^2}\right)^n}{c_2\left(\frac{R}{m^2}\right)^n+1},
\end{equation}
where $n$, $c_1$ and $c_2$ are positive dimensionless parameters and $m^2\equiv8\pi G\bar{\rho}_{c,0}/3=H_0^2\Omega_m$ is a mass scale introduced in the model. The new dynamical degree of freedom is therefore represented by the scalar field (commonly called scalaron), $f_R \equiv \mathrm{d} f(R)/\mathrm{d}R$, that can be approximated to
\begin{equation}\label{eq:scalaron}
f_R \approx-n \frac{c_1}{c_2^2}\left(\frac{m^2}{R}\right)^{n+1}=\left(\frac{\Omega_m+4\Omega_\Lambda}{\Omega_m a^{-3}+4\Omega_\Lambda}\right)^{n+1}f_{R0},
\end{equation}
with $f_{R0}$ being the dimensionless scalar field at present time. The rightmost expression of Eq.~\eqref{eq:scalaron} has been tuned to mimic to first order the background expansion of the \lcdm\ model by setting $c_1/c_2=6\Omega_\Lambda/\Omega_m$. Moreover, the background curvature is given by $R=6(2H^2+\dot{H})$, so that the bound condition $f(R)\rightarrow-2\Lambda$ is satisfied as consequence of requiring equivalence with \lcdm\ when $\left|f_{R0}\right|\rightarrow0$. If the exponent $n$ is assigned a specific value, then the model model is fully specified by only one free parameter, $f_{R0}$. In the simulations conducted for this paper, we explored two values of the exponent $n$: the first being $n=1$, where we systematically vary $\left|f_{R0}\right|$ as $10^{-4},\ 10^{-5}$ and $10^{-6}$, denoted as F$4_1$, F$5_1$ and F$6_1$, respectively. The second value is $n=2$, where we vary $\left|f_{R0}\right|$ as $10^{-3.5},\ 10^{-5}$ and $10^{-6.5}$, labelled as F$3.5_2$, F$5_2$, and F$6.5_2$. These choices were made to assess the robustness of our technique and to comprehensively cover current observational constraints.
\begin{figure}
   \centering
   \includegraphics[width=\linewidth]{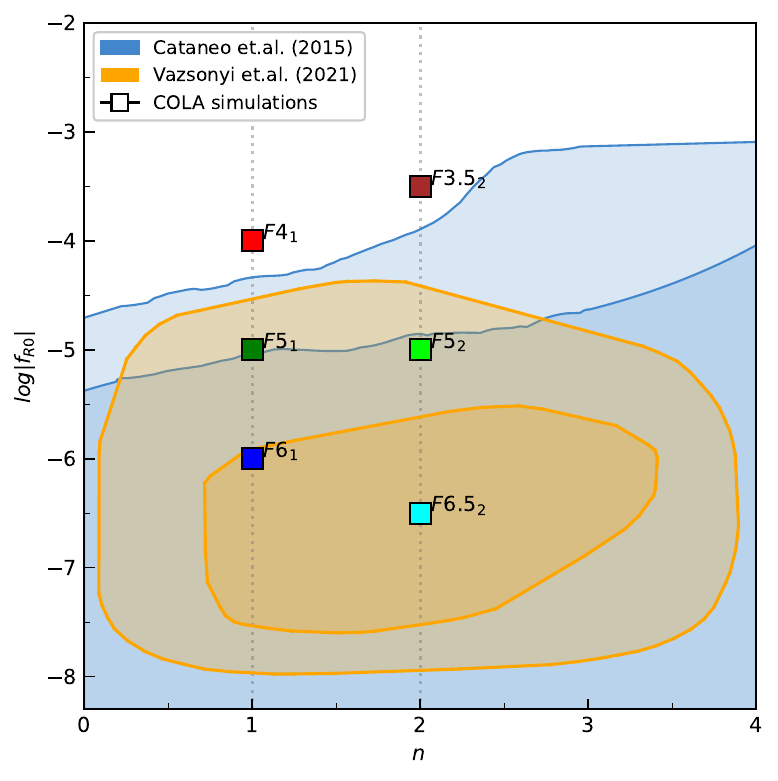}
      \caption{
      \small{Schematic representation of HS $f(R)$ constraints in the $f_{R0}-n$ plane. The location of the colored squares show the location of our simulations. The actual constraints at $1-$ and $2-\sigma$ confidence levels are taken from \cite{Cataneo_fR_constraints_2015} (blue contours) and \cite{Vazsonyi_fR_constraints_2021} (orange contours).}}
         \label{fig:ConstraintsCOLASims}
\end{figure}

Current constraints on HS models set the upper limit on the scalaron field at present time $|f_{R0}|\leqslant 5.68\times10^{-7}$ at $2\sigma$ confidence levels when using cluster abundances and galaxy clustering \citep{Liu_fR_Constraints_2021} and $f_{R0}\leqslant\times10^{-6.75}$ when combining CMB, BAO, SNIa, cosmic chronometers and redshift-space distortions \citep{Wang_PLack_fR_constraints_2021}. Previous analysis were significantly relaxed constraining $|f_{R0}|<10^{-4.79}$ \citep{Cataneo_fR_constraints_2015}, $<5\times10^{-6}$ \citep{Shirasaki_fR_constraints_2016} and $<3.7\times10^{-6}$ at $2\sigma$ confidence level \citep{Boubekeur_fR_constraints_2014}. Gravitational wave detection has also provided strong bounds, $|f_{R0}|<5 \times 10^{-7}$ at $1\sigma$ confidence \citep{Vainio_fR_constraints_2017} and the so-called ``galaxy clustering ratio'' $|f_{R0}|<5\times10^{-6}$ at $68\%$ confidence\citep{Bel_fR_constraints_2015}. This is competitive with astrophysical tests \citep{Jain_fR_constraints_2013} and dwarf galaxies analysis \citep{Vikram_fR_constraints_2013} which provide tighter narrow the constraints to $|f_{R0}|\leq5\times10^{-7}$. The latest bounds using Subaru Hyper Suprime-Cam year 1 data are in agreement with $\log f_{R0} = -6.38^{+0.94}_{-1.41}$ and $n=1.8^{+1.1}_{ -1.5}$, representing substantial improvement in the constraints \citep{Vazsonyi_fR_constraints_2021}. Envisioning the future data, it is expected that Euclid's observations would constrain $\log|f_{R0}|$ at the 1\% level using the full combination of spectroscopic and photometric galaxy clustering \citep{Casas_2023arXiv230611053C}. Fig. \ref{fig:ConstraintsCOLASims} depicts the constraints on the HS model in the $\log|f_{R0}|-n$ plane. The orange confidence regions are reproduced from \citet{Cataneo_fR_constraints_2015} and correspond to the 68.3\% and 95.4\% confidence levels from the combination of clusters, CMB (Planck + WMAP + lensing + ACT and SPT) and SNIa + BAO. The blue confidence regions are reproduced from \citet{Vazsonyi_fR_constraints_2021} from cosmic shear analysis using Subaru Hyper Suprime-Cam year 1 data \citep{2019PASJ...71...43H}. The colored squares in Fig. \ref{fig:ConstraintsCOLASims} correspond to the location of the MG simulations in the $\log|f_{R0}|-n$ plane relative to the actual constraints of these parameters. Therefore, our simulations spans models that faithfully reproduce nearly all features of the standard model to those that are ruled out by astrophysical probes.

An interesting feature of the HS model is that evades stringent constraints of deviations of GR on the Solar system scale by means of the Chameleon mechanism \citep{Khoury_fR_chameleon2004_I,Khoury_fR_chameleon2004_II,MotaShaw_2007PhRvD}. This mechanism leads to a complex interplay between the matter distribution and the magnitude of the fifth force that hide (or screen) any modification of gravity in some regions and separations. In fact, the scalar field's action generates intriguing effects across all scales, with its influence effectively concealed in local environments \citep{2012ApJ...756..166W,2014PhRvD..89h4023L,2016A&A...595A..40I}. According to local gravity constraints, the fifth force has an extremely weak strength. However, high-density environments may conceal the fifth force via the Chameleon mechanism. This is attributed to the inherent environmental dependence induced by the properties of dark matter distributions in the Chameleon mechanism, which operates in such a way that the scalar field acquires a substantial mass in denser environments, rendering the fifth force negligible. Conversely, on cosmological scales, the scalar field remains light, resulting in significant modifications to gravity \citep{Khoury_fR_chameleon2004_I,Khoury_fR_chameleon2004_II,Will_screening_2014LRR}. A detailed discussion of the clustering in the HS model, based on simulations of halos and galaxies both in real- and redshift-space, can be found in \citep{Garfa2019-MGMnu,Aguayo2019,2019JCAP...06..040W,2021PhRvD.103j3524G}.

The structure formation driven by the HS $f(R)$ scalar field amplifies the growth of structure, leading to higher-density peaks compared to \lcdm. Conversely, on small scales, the modifications tend to decrease structure formation, resulting in a smoother density field. This effect can be seen in the distribution of halos since it impacts their abundance and thereby induces a enhancement or suppression relative to the \lcdm\ scenario. In the linear regime, the perturbations are described by the modified Poisson equation (in Fourier space) for the gravitational potential $\Phi$ as $k^2 \Phi=-4 \pi G_{\mathrm{eff}} a^2 \delta \rho$ with $\delta\rho$ being the fluctuation around the mean density and $G_\mathrm{eff}$ is the effective Newton's constant that takes the form \citep{Tsujikawa_Geff_2008, Esposito_Geff_2001}
\begin{equation}\label{eq:effgravconst}
\frac{G_{\mathrm{eff}}}{G_\mathrm{N}}=\left\{\begin{array}{lc}
1 & \mathrm{GR} \\
1+k^2 /\left[3\left(k^2+a^2 m_{f_R}^2\right)\right] & f(R) .
\end{array}\right.
\end{equation}
Here, $m_{f_R}$ is the mass of the scalar fluctuations (not to be confused with the mass scale $m$ of the HS model). This quantity plays a crucial role in the Chameleon mechanism as it establishes when the scalar field is suppressed. The linear growth for the matter fluctuations in these models is governed by \citep{Lombriser_GeffD_2014}:
\begin{equation}\label{eq:growthfactor}
D^{\prime \prime}+\left[2-\frac{3}{2} \Omega_{\mathrm{m}}(a)\right] D^{\prime}-\frac{3}{2} \frac{G_{\text {eff }}}{G_N} \Omega_{\mathrm{m}}(a) D\approx0,
\end{equation}
where $D=D(a,k)$ is the linear growth function and the derivatives are with respect to $\ln a$. Note that both, $G_{\mathrm{eff}}$ and $D$ are functions of time- and scale-dependent, unlike the GR case in which both are independent of the scale.
\begin{figure}
    \centering
    \includegraphics[width=\linewidth]{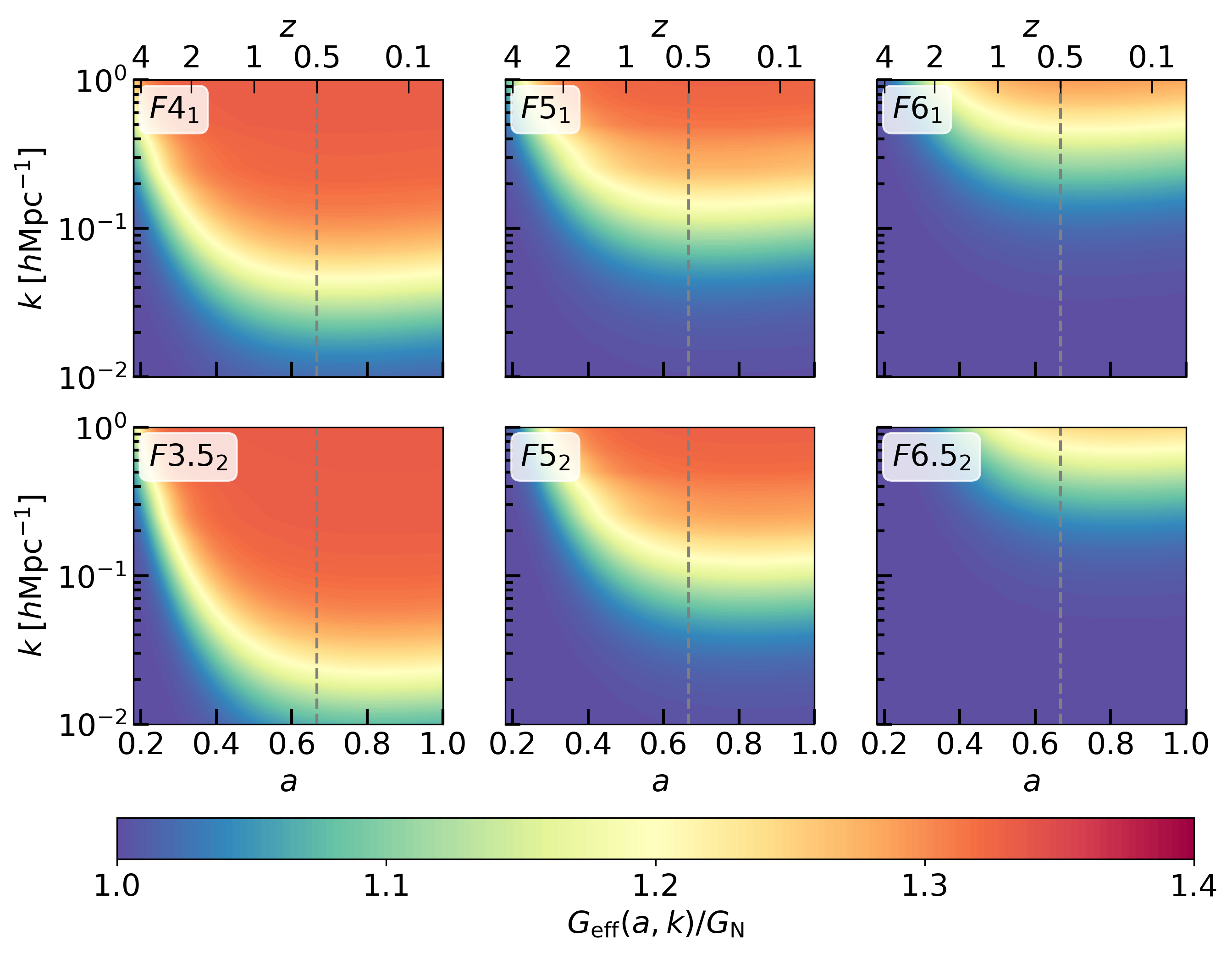}
    \caption{\small{Normalised effective gravitational constant of the HS model variants as labelled in the respective panels. Color intensity represents variations of $G_\mathrm{eff}(k,a)$ across the scale factor $a$ and Fourier modes $k$. The dashed gray line corresponds to the redshift of interest in our simulations, \ie\ $z=0.5$. The color bar indicates the intensity of deviation from the \lcdm\ model.}}
    \label{fig:Geff2D}
\end{figure}
\begin{figure}
    \centering
    \includegraphics[width=\linewidth]{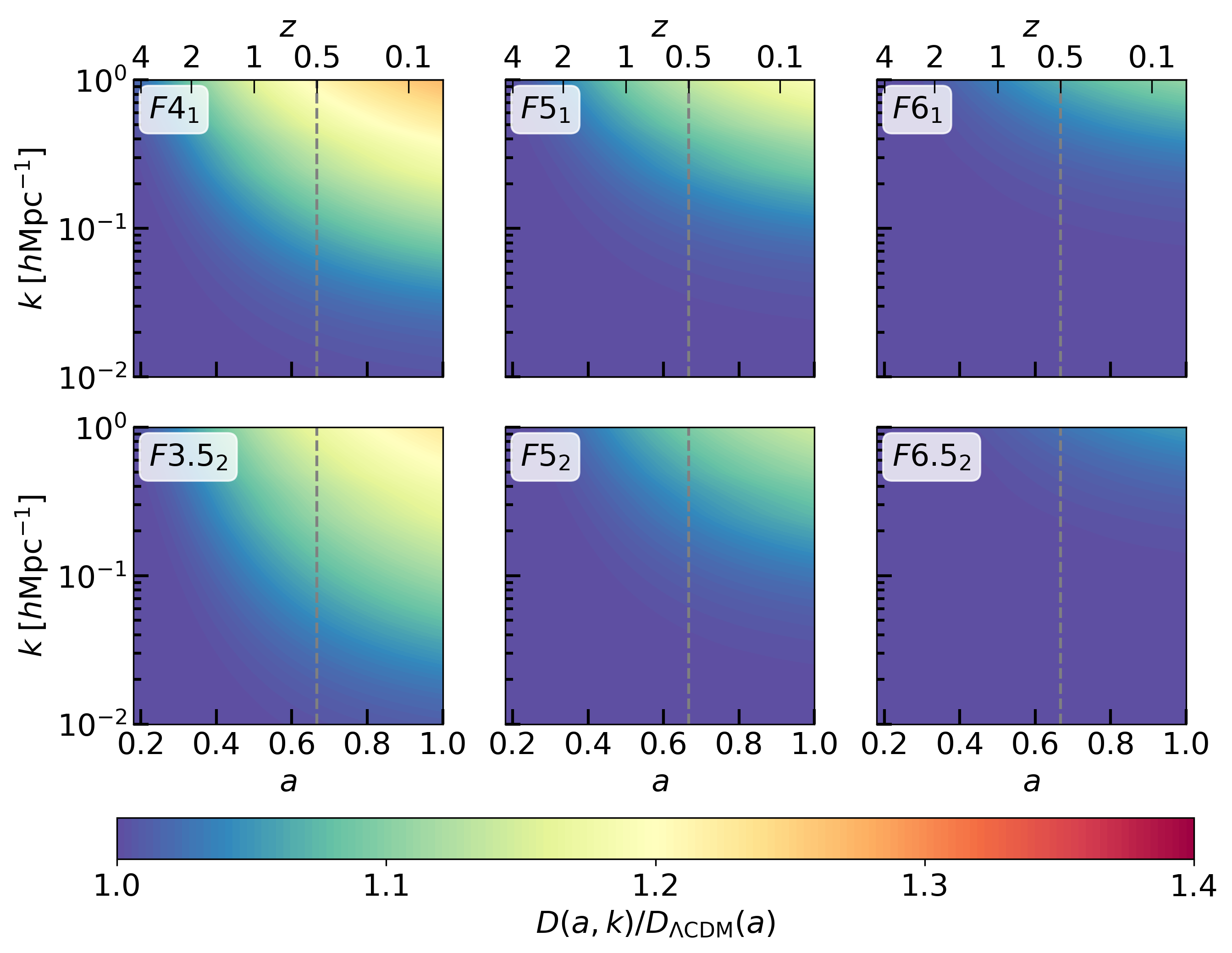}
    \caption{\small{Similar to Fig. \ref{fig:Geff2D}, but now for the linear growth factors of the MG models.}}
    \label{fig:Growth2D}
\end{figure}
Figs. \ref{fig:Geff2D} and \ref{fig:Growth2D} display the numerical solution of the effective gravitational constant and the linear growth factor as a function of time and scale, eqs. \eqref{eq:effgravconst} and \eqref{eq:growthfactor} respectively. The different panels of the figures refer to the different HS models considered in this study, as indicated by their labels. The dashed gray line at $z=0.5$ ($a=0.\bar{6}$), represents the redshift of the simulation snapshots used for our analysis. $G_\mathrm{eff}$ varies with the inverse square of the time and scale as can be appreciated by the contours of Fig. \ref{fig:Geff2D}. The area above the contour defined by $G_\mathrm{eff}(k,a)=1$ shrinks with the scalar field strength $f_{R0}$, capturing deviations ranging from 5\% up to 40\% in the most extreme case (model $F3.5_2$). The intensity of the normalised $G_\mathrm{eff}$ decreases as $f_{R0}$ approaches zero, which is the \lcdm\ scenario. This trend is clearly followed by the models $F5_1$, $F6_1$ and $F6.5_2$ respectively. Similarly, in Fig. \ref{fig:Growth2D} we observe the variation of $D(a,k)$ across different scales and times. The disparity in the growth factor can reach up to 30\% for the plotted time interval and scales, \ie, $k\in[10^{-2}, 1]$ \hMpc\ and $a\in[0.2, 1]$. The most significant deviations from the \lcdm\ growth factor can be seen at high Fourier modes, $k \sim 1$ \hMpc\, meaning middle non-linear regime of the structure formation, and at low redshifts. For models with larger $|f_{R0}|$, such as $|f_{R0}| = 10^{-4}$, the chameleon screening is inefficient, as illustrated by the significant discrepancies in $D$ of $F4_1$ compared to \lcdm.
\begin{figure}
    \centering
    \includegraphics[width=\linewidth]{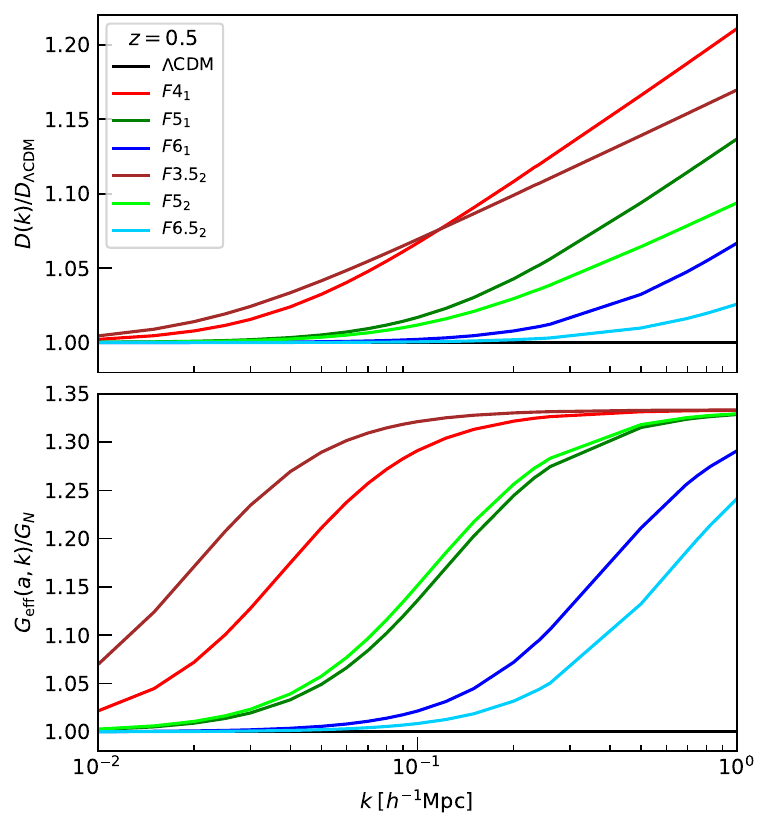}
    \caption{\small{Comparison of the normalised linear growth factors (upper panel) and effective gravitational constant (lower panel), of the MG models as labelled. The plot display the variations of these quantities as a function of $k$ for the interested redshift of the simulations $z=0.5$.}}
    \label{fig:Geff_Growth1D}
\end{figure}
Fig. \ref{fig:Geff_Growth1D} shows the dynamics of $G_\mathrm{eff}$ and $D$ for the redshift of interest, $z=0.5$; this essentially represents the cross-section of the contour plots depicted in Figs. \ref{fig:Geff2D} and \ref{fig:Growth2D} at the given redshift. Here, we clearly discern the deviations of each HS model from \lcdm. At this redshift, $G_\mathrm{eff}$ exhibits deviations of up to 33\% for the most extreme model, $F4_1$ (red line), and deviations of up to 20\% for the closest model to \lcdm, \ie, $F6.5_2$ (cyan line). Intermediate deviations are observed for the remaining HS models, as depicted in the figure. Conversely, the growth factor follows a monotonic trend with respect to the scale and $|f_{R0}|$, with deviations reaching 40\% for $F4_1$ and remaining below 3\% for $F6.5_2$. However, at small wavenumbers, $D$ becomes increasingly indistinguishable from \lcdm, especially for models with the lowest $|f_{R0}|$. In contrast, $G_\mathrm{eff}$, can still be distinguished from \lcdm\ up to 7\% at these scales for models with high $|f_{R0}|$.

Overall, the enhancement in the growth factor can be attributed to the effective mass of the scalar field, which limits the range of interaction of the fifth force, leaving, as a consequence, the growth on larger scales almost unaffected. In terms of distances, the effective length at which modifications of the gravitational potential take effect is given by the Compton wavelength of the scalaron field, which in turn can be expressed in terms of $f_{R0}$, given by:
\begin{equation}\label{eq:lambdaCompton}
\lambda_C\equiv\frac{1}{m_{f_R}}\approx\frac{2997.92}{a}\sqrt{\frac{(n+1)\left|f_{R 0}\right|\left(4-3\Omega_m\right)^{n+1}}{\left[\Omega_m\left(a^{-3}-4\right)+4\right]^{n+2}}}~~h^{-1}\mathrm{Mpc}.
\end{equation}
In addition to the scale-dependent effect due to the screening mechanism, the Compton wavelength sets a cut-off where structures cluster according to GR (distances greater than $\lambda_C$). Below this distance, the growth rate increases being modulated by the MG dynamics, giving rise to specific scale-dependent patterns. On large scales, $\lambda_Ck/a \ll 1 $ the perturbation equation is identical to that in GR, however, on smaller scales, $\lambda_Ck/a\gg1$, gravity is enhanced by a maximal factor of 4/3 \citep{Pogosian_growth_fR_2008, Huillier_MG_Halos2017}.

In the upcoming sections, we present the simulations in detail, followed by an overview of the \texttt{BAM} method and the calibration analysis.

\section{MG simulations and training data set}\label{sec:sims}
We run a set of high-resolution \texttt{COLA} simulations of \lcdm\ and six HS models with $|f_{R0}|$ consistent with current constraints (see \textsection\ref{sec:mgmodel}). The simulations were performed with the \texttt{COLA} Solver implemented in the publicly available FML library\footnote{\url{https://github.com/HAWinther/FML}}, which succeeded \texttt{MG-PICOLA}. The FML-\texttt{COLA} solver extends the \texttt{COLA} method for simulating cosmological structure formation from \lcdm\ to theories with scale-dependent growth such the HS model. It also includes a fast approximate screening method described by \citet{2015PhRvD..91l3507W}. The simulated HS models correspond to six combinations of the exponent of the modified gravity function, $n$, and the magnitude of the scalar field $|f_{R0}|$. These models correspond to the pairs $(|f_{R0}|,\,n)\in\{(10^{-4},\,1),\:(10^{-5},\,1),\:(10^{-6},\,1),\:(10^{-3.5},\,2),\:(10^{-5},\,2),\:(10^{-6.5},\,2)\}$ which are denoted as $\{F4_1,\, F5_1,\,F6_1,\,F3.5_2,\,F5_2,\,F6.5_2\}$ respectively. The simulations are consistent with the best-fit parameters of Planck 2018 cosmology \citep{Planck_parameters2020}, and feature the dynamics of $2048^3$ dark matter particles in a comoving box of $1\,Gpc/h^3$ on a side. The simulations commenced at $z=99$, with initial conditions generated via a modified version of the 2LPTic code \citep{Scoccimarro_1998MNRAS,Crocce_2006MNRAS} implemented in FML. The HS models considered in this work share identical initial conditions with the \lcdm\ run, as deviations in gravity are not expected at such high redshift. The evolution of the DM particles extends up to redshift $z=0.5$, comprising 100 time-steps, which is a fairly large number. This finer temporal sampling allows more detailed tracking of the evolution of the particle distribution, potentially capturing more subtle effects in the simulation, albeit at the cost of increased computational resources. 
\begin{table}
\caption{\small{Cosmological parameters employed in the COLA simulations along with key features of the simulation setup.}}
\label{table:COLAsetup}
\centering
\begin{tabular}{ccccc}
\cline{1-2} \cline{4-5}
\multicolumn{2}{c}{\textbf{Fid. cosmology}}                                                                      &  & \multicolumn{2}{c}{\textbf{Simulation setup}}                                   \\ \cline{1-2} \cline{4-5}
\cellcolor[HTML]{EFEFEF}{\color[HTML]{333333} $\Omega_m$} & \cellcolor[HTML]{EFEFEF}{\color[HTML]{333333} 0.311} &  & \cellcolor[HTML]{EFEFEF}Boxsize    & \cellcolor[HTML]{EFEFEF}$1\;h^{-1}$Gpc           \\
$\Omega_{cdm}$                                            & 0.2621                                               &  & $N_p$                              & $2048^3$                                   \\
\cellcolor[HTML]{EFEFEF}$\Omega_b$                        & \cellcolor[HTML]{EFEFEF}0.0489                       &  & \cellcolor[HTML]{EFEFEF}Grid force & \cellcolor[HTML]{EFEFEF}$2048^3$           \\
$h$                                                       & 0.6766                                               &  & $M_p$                              & $1.005\times10^{10}~M_\odot/h$             \\
\cellcolor[HTML]{EFEFEF}$n_s$                             & \cellcolor[HTML]{EFEFEF}0.9665                       &  & \cellcolor[HTML]{EFEFEF}IC         & \cellcolor[HTML]{EFEFEF}2LPT $z_{ini}=99$ \\
$A_s$                                                     & 2.105e-09                                            &  & Steps                              & 100                                        \\
\cellcolor[HTML]{EFEFEF}$\sigma_8$                        & \cellcolor[HTML]{EFEFEF}0.8102                       &  & \cellcolor[HTML]{EFEFEF}$k_{\mathrm{Ny}}$   & \cellcolor[HTML]{EFEFEF}$6.43$ \hMpc       \\ \cline{1-2} \cline{4-5}
\end{tabular}
\end{table}
\begin{table}[]
\caption{\small{Normalised effective gravitational constant and linear growth factors of each variant of the HS model, evaluated at their respective Compton length and simulation redshift $z=0.5$.}}
\label{table:MGdescription}
\begin{adjustbox}{width=\linewidth,center}
\begin{tabular}{llccccc}
\hline
\multicolumn{1}{c}{\textbf{Model}} & $|f_{R0}|$  & $n$ & \begin{tabular}[c]{@{}c@{}}$\lambda_\mathrm{C}|_{z=0.5}$ \\ {[}Mpc/h{]}\end{tabular} & \begin{tabular}[c]{@{}c@{}}$k_\mathrm{C}|_{z=0.5}$ \\ {[}h/Mpc{]}\end{tabular} & $\frac{G_{\mathrm{eff}}}{G_\mathrm{N}}|_{z=0.5\,k=k_\mathrm{C}}$ & $\frac{D_{\mathrm{MG}}}{D_{\Lambda\mathrm{CDM}}}|_{z=0.5\,k=k_\mathrm{C}}$ \\ \hline
\rowcolor[HTML]{EFEFEF} 
GR                                 & -           & -   & -                                                                                    & -                                                                              & 1.00                                                         & 1.00                                                                   \\
$F4_1$                    & $10^{-4}$   & 1   & 26.20                                                                                & 0.24                                                                           & 1.33                                                         & 1.08                                                                   \\
\rowcolor[HTML]{EFEFEF} 
$F5_1$                    & $10^{-5}$   & 1   & 8.29                                                                                 & 0.76                                                                           & 1.33                                                         & 1.05                                                                   \\
$F6_1$                    & $10^{-6}$   & 1   & 2.62                                                                                 & 2.40                                                                           & 1.33                                                         & 1.05                                                                   \\
\rowcolor[HTML]{EFEFEF} 
$F3.5_2$                  & $10^{-3.5}$ & 2   & 51.23                                                                                & 0.12                                                                           & 1.33                                                         & 1.16                                                                   \\
$F5_2$                    & $10^{-5}$   & 2   & 9.11                                                                                 & 0.69                                                                           & 1.33                                                         & 1.13                                                                   \\
\rowcolor[HTML]{EFEFEF} 
$F6.5_2$                  & $10^{-6.6}$ & 2   & 1.62                                                                                 & 3.88                                                                           & 1.33                                                         & 1.08                                                                   \\ \hline
\end{tabular}
\end{adjustbox}
\end{table}

Our setup ensures a mass resolution of $M_{p}=1.005\times10^{10}$ \Msunh\ and a Nyquist frequency given by $k_{\mathrm{Ny}}=6.43$ \hMpc. The table \ref{table:COLAsetup} shows the key base cosmological parameters employed (left side) as well as the general setup of the \texttt{COLA} solver (right side). The table \ref{table:MGdescription} presents the normalised effective gravitational constant and linear growth factors for various variants of the HS model at a redshift of $z=0.5$. The Compton length and the corresponding wavenumber are provided for each model, indicating the characteristic scale at which the modifications to gravity become significant. As expected, the $G_{\mathrm{eff}}|_{z=0.5\,k=k_\mathrm{C}}$) remains approximately constant across all models with a value of $1.33\sim4/3$ at the characteristic scale. Similarly, the table illustrates that at the Compton scale, the impact of modified gravity on the linear growth factor ratio is relatively small, with deviations from \lcdm\ reaching up to 16\%.
\begin{figure*}
    \centering
    \includegraphics[width=\textwidth]{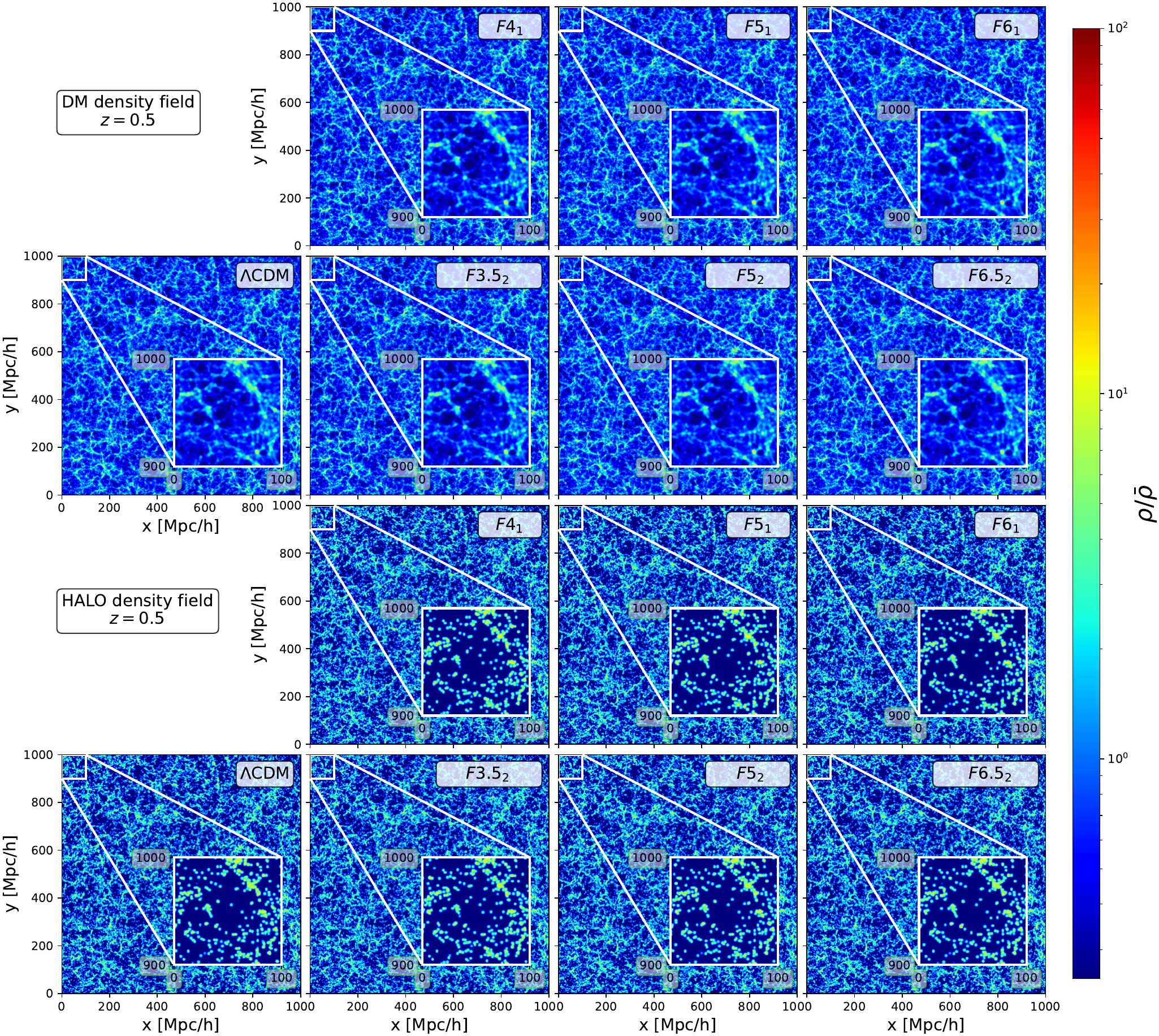}
    \caption{\small{Projected density field of DM and halos in a region of 1000$\times$1000$\times$20 \Mpch\ from seven different cosmologies at redshift $z=0.5$. The upper panels display the DM density field, while the lower panels exhibit the corresponding halo density field for the MG models as labelled. These maps highlight the remarkable similarity in the density fields of MG and \lcdm\ models. They also show that some models give rise to the formation of pronounced filaments, characterised by either substantial thickness or length, as well as an increased prevalence of halos.}}
    \label{fig:density_fields_grid_reduced}
\end{figure*}

DM halos were identified using the Friends-of-Friends (FoF hereafter) algorithm, which connects particles within a distance less than a specified linking length, $b=0.2$, measured in units relative to the mean inter-particle distance \citep{1985ApJ...292..371D}. The linking length value has been shown to be valid for \texttt{COLA} simulations as already employed in previous works \citep[see \eg][]{2015MNRAS.449..848H,2016MNRAS.459.2118K,2016MNRAS.459.2327I,Ferrero_icecola_mocks_2021}. We opt for the FoF halo finder over \texttt{ROCKSTAR} due to the poor performance of the later one when using \texttt{COLA} simulations \citep{2021JCAP...09..021F}. This finding is widely discussed by \citet{2021JCAP...09..021F}, highlighting that the default \texttt{ROCKSTAR} settings can lead to statistical discrepancies in the halo properties when compared to $N$-body simulations, with differences of up to 25\% in the halo mass function and approximately 10\% variations in the power spectrum at $k\sim0.4$ \hMpc. 
The reference catalogues used in this study refer to the halo distribution at $z=0.5$ of each MG model plus \lcdm\ as benchmark model. We have chosen halos that contain no fewer than 83 DM particles, corresponding to an average mass-cut of FoF masses of $M_\mathrm{FoF}\gtrsim 8.3\times10^{11}$ \Msunh, which is nearly is represented by $\sim3\times10^{6}$ distinct halos per catalogue. This criterion agrees with the mass-cut employed in previous researches to assess the performance of bias mapping methods \citep{2017MNRAS.472.4144V,2019MNRAS.483L..58B}, as well as mass-scale at which DM halos can host Emission Line Galaxies (ELGs) in the extended Baryon Oscillation Spectroscopic Survey (eBOSS) \citep[for details see][]{Alam_2020MNRAS}.
DM fields are obtained from the DM particle distribution applying a cloud-in-cell (CIC) mass assignment scheme onto a $N^3 =256^3$ mesh (equivalent to a resolution of $3.9$ \Mpch\ per cell). Halo number counts are obtained using a nearest-grid-point (NGP) scheme \citep{1981csup.book.....H} using the same resolution as for the DM catalogues.

Figure \ref{fig:density_fields_grid_reduced} depicts the density field at $z=0.5$ of DM particles (upper panels) and halos (bottom panels) in a region of 1000$\times$1000$\times$20 \Mpch\ for the different gravity models considered in this work. The color bar on the right displays the magnitude of the density perturbations. The inner zoom-panels highlight the similar web-like structures across these models, despite the presence of modified gravity effects. However, some differences are visible in the halo distribution, showing-up regions with a higher density of halos than in \lcdm. These discrepancies stem from modifications in gravity, which amplify gravitational forces on small scales, thereby impacting the formation of prominent filaments, distinguished by either significant thickness or considerable length, along with a heightened abundance of halos. Such enhancements translate into a more pronounced abundance of rare and massive halos in the nonlinear regime, as illustrated by cosmological simulations \citep{2009PhRvD..80h3505S}. 

\section{\texttt{BAM} approach and calibration analysis}\label{sec:app}

In this section, we provide an overview of the \texttt{BAM} method used create mock catalogues, along with the calibration conducted with the MG halo catalogues. Lastly, we present the summary statistics performance of the mapping for the different cosmologies.
We refer the reader to \citet[][]{2023A&A...673A.130B} for more details on the method.
\subsection{Calibration: kernel and halo bias}\label{sec:BAMcalibrations}
\texttt{BAM} is a non-parametric method designed to reproduce the halo number counts, $N_{h}$, of a reference catalogue within a mesh, using a target DM density field (TDMF hereafter). The idea behind the method consists of mapping the DM halo distribution (DMH) by exploiting the concept of stochastic bias \citep{1999ApJ...520...24D,2001MNRAS.320..289S,2002MNRAS.333..730C}, minimisation of a cost function based on the power spectrum of target variables and computing an iterative kernel that corrects for missing power towards small scales. The method has the capability of capturing the different properties of the halo bias by assuming that the number counts of halos in a volume cell depends on a set of properties of the DM density field evaluated in the same cells of volume $\partial V$.  In this context, the bias is represented as a conditional probability distribution (\ie, the CPD) of the halo number counts, obtained directly from the reference simulation as a multidimensional histogram, $\mathcal{B}\left(N_{\mathrm{h}} \mid \Theta_{\mathrm{dm}}\right)_{\partial V}$. The TDMF field can be either the one obtained from the full $N$-body snapshot downgraded to a mesh of $N^3$ grid cells, or it can be generated by an approximate gravity solver that evolves the downgraded initial conditions of the $N$-body simulations in a mesh of the same dimensionality. 

The calibration process, wherein the halo bias and a so-called \texttt{BAM} kernel are obtained solely from the two-point statistics of the reference catalogue as a target, involves an iterative procedure with Markov Chain Monte Carlo rejection algorithm. The aim of the \texttt{BAM} kernel is to adjust the DM density field through a convolution to match the reference tracer power spectrum. This process achieves approximately 1\% accuracy in the power spectrum up to the Nyquist frequency, as demonstrated in previous studies \citep[see \eg,][]{2019MNRAS.483L..58B,2020MNRAS.493..586P,Balaguera_2020MNRAS,2022MNRAS.512.2245K,2023A&A...673A.130B}. The algorithm accounts for cross-correlations and other dependencies such local and non-local properties like density and cosmic web type. We include these in the form of a cosmic-web classification (\ie\, knots, filaments, sheets and voids), which are obtained via the eigenvalues of the tidal field \citep[see e.g.][]{2007MNRAS.375..489H,2009MNRAS.396.1815F}. The outputs of this procedure (halo-bias and kernel) enable the generation of new halo samples with the same probability density function (PDF) as the reference catalogue, representing the desired halo distribution. This mapping technique can then be applied to generate accurate halo distributions across various initial conditions while preserving the background cosmology.

\begin{figure}
    \centering
    \includegraphics[width=\columnwidth]{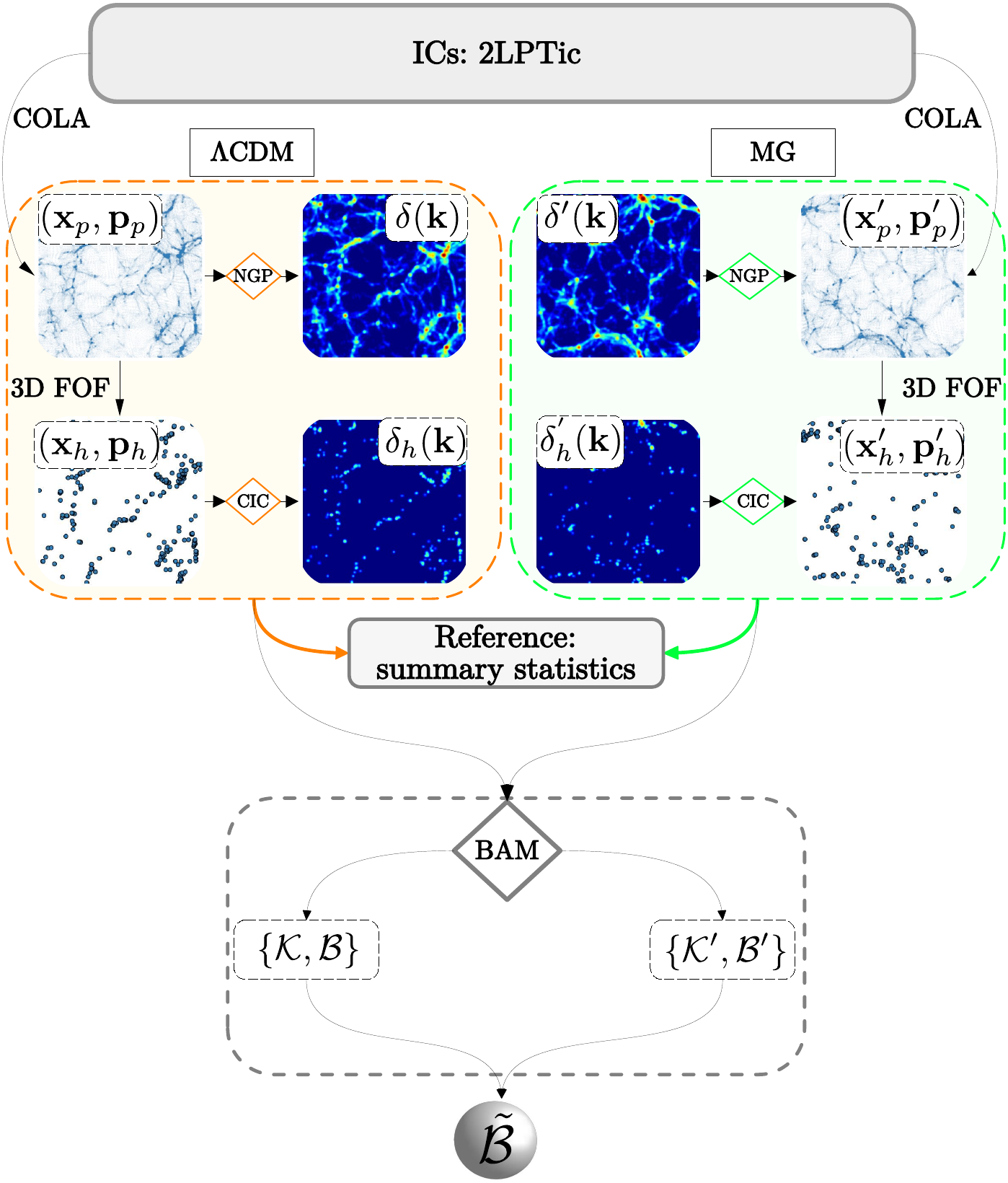}
    \caption{\small{Flowchart depicting the kernel and bias calibration using the \texttt{COLA} simulations as a reference to subsequently obtain the mapping bias relation $\tilde{\mathcal{B}}$.}
    }
    \label{fig:Flowchart}
\end{figure}

\subsection{Mapping MG cosmologies} 

We explore two calibration experiments to assess the performance of \texttt{BAM} in mapping MG cosmologies. The first one consists on employing the MG DM fields to generate the corresponding MG halo number counts; this calibration is henceforth referred as consistent-field calibration and can be represented by the following operation: $\delta_{\mathrm{DM}}^{\mathrm{MG}}$\SymbolBiasEffective$N_h^{\mathrm{MG}}$, where the operator \SymbolBiasEffective\ indicates that the halo-bias relation from \texttt{BAM} has been applied to produce the mock catalogue. The second experiment, referred to as cross-field calibration from now on, consists on utilising the DM field of the \lcdm\ model to generate the six different halo number counts of the MG HS models. The second calibration, represented by the transformation $\delta_{\mathrm{DM}}^{\text{\lcdm}}$\SymbolBiasEffective$N_h^{\mathrm{MG}}$, is of interest because it allow us to create fast mocks of MG cosmologies without the need to run the corresponding MG simulations, which are usually computationally more expensive than the \lcdm\ ones. 

In terms of the effective field theory (EFT) of LSS, which in turn is based on cosmological perturbation theory, the transformations describing our two calibrations can be expressed as a linear superposition of fields, $O$, with corresponding bias coefficients $b_O$ ($\tilde{b}_O$) as well as stochastic contributions given by a field $\epsilon$ ($\tilde{\epsilon}$) and coefficients $c_{\epsilon,O}$ ($\tilde{c}_{\tilde{\epsilon},O}$) \citep[see \eg,][]{2016PhRvD..93f3512S,2018PhR...733....1D}. That is, the consistent-field calibration can be expressed as
{\small
\begin{equation}\nonumber
    \delta^{\mathrm{MG}}_h(\boldsymbol{x}, \tau) = \sum_O\left[b_O(\tau)+c_{\epsilon, O}(\tau) \epsilon(\boldsymbol{x}, \tau)\right] O\left[\delta^{\mathrm{MG}}_{\mathrm{DM}}\right](\boldsymbol{x}, \tau)+\epsilon(\boldsymbol{x}, \tau), 
\end{equation}}
while for the cross-field calibration
{\small
\begin{equation}\nonumber 
    \delta^{\mathrm{MG}}_h(\boldsymbol{x}, \tau) = \sum_O\left[\tilde{b}_O(\tau)+\tilde{c}_{\tilde{\epsilon}, O}(\tau) \tilde{\epsilon}(\boldsymbol{x}, \tau)\right] O\left[\delta^{\text{\lcdm}}_{\mathrm{DM}}\right](\boldsymbol{x}, \tau)+\tilde{\epsilon}(\boldsymbol{x}, \tau),
\end{equation}}
with $\boldsymbol{x}$ representing the position in space and $\tau$ being the time variable of the galaxy bias renormalisation group. The tilde over the parameters in these expressions indicates their association with the cross-field calibration, distinguishing them from those in the consistent-field calibration.

\texttt{BAM} performs an iterative process aimed at mapping the DM field to the reference halos by minimising the 2-point statistics as previously discussed in \S\ref{sec:BAMcalibrations} \footnote{During the calibration process, \texttt{BAM} fixes the number counts to those of reference catalogues. That is, the target counts of the MG halos reproduce its PDF by construction}. Fig. \ref{fig:Flowchart} illustrates the different steps of the calibration process through a flowchart. Read from top to bottom, the process starts with the creation of initial conditions using 2LCTic, followed by the evolution of the DM particles with the \texttt{COLA} gravity solver until obtain its distribution at $z=0.5$. Then, the FoF algorithm is applied to identify halo structures, whose number counts on the mesh are used as reference. The two branches of Fig. \ref{fig:Flowchart} distinguish between the benchmark \lcdm\ model and MG models. The DM density fields and halo number counts are used as input in the \texttt{BAM} algorithm to perform the iterative process to produce a calibrated mock catalogue. The process ends with obtaining the \texttt{BAM}-kernel $\mathcal{K}$ ($\tilde{\mathcal{K}}$) as well as the corresponding bias relationship, $\mathcal{B}$ ($\tilde{\mathcal{B}}$), depending on the input DM density field, either \lcdm\ or MG. Once the respective calibrations are obtained, we compute and compare the summary statistics, including density fields, PDFs, power spectra and reduced bispectra \footnote{We used the code \texttt{bispect} \url{https://github.com/cheng-zhao/bispec} }.

\begin{figure}
    \centering
    \includegraphics[width=\linewidth]{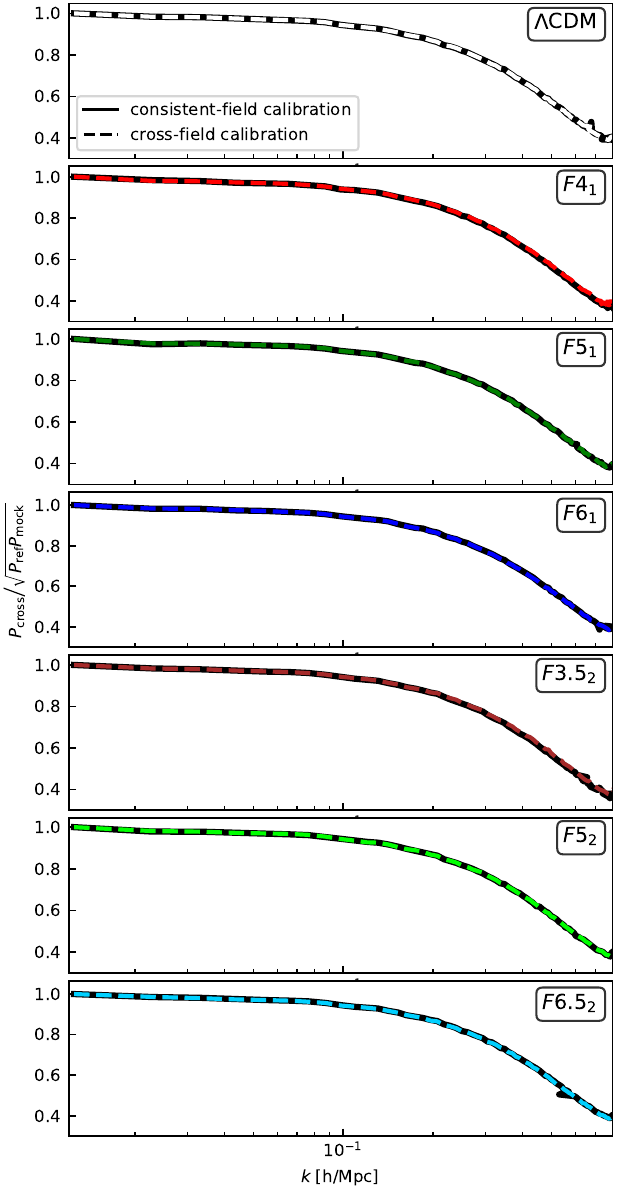}   
    \caption{\small{Comparison of the cross power spectrum between the mock catalogues generated by BAM and their respective reference catalogues for the two calibrations: consistent-field calibration (solid line) and cross-field calibration (dashed line). The color of the dashed lines has been kept for consistency with the other plots.}}
    \label{fig:crosspk}
\end{figure}
\subsection{Performance and summary statistics}\label{subsec:performance_statistics}
In this section, we present our main findings in terms of the summary statistics.
Mapping the DM density fields with \texttt{BAM} provides the corresponding number counts of halos for each of the calibrations studied. Consequently, for each MG model we generate two mock catalogues: one for the consistent-field calibration, denoted as $N^{\Vert}_h$, and another for the cross-field calibration, denoted as $N^{\times}_h$. The respective reference halo catalogue is labelled $N^{\mathrm{ref}}_h$. Firstly, we look into the cross-power spectra between the references and the calibrated fields. Fig. \ref{fig:crosspk} shows the cross power spectra of both calibration types across the MG cosmologies and \lcdm. As seen in the figure, there are no significant fluctuations between the density fields in either calibration. In fact, the correlation exceeds 75\% up to scales of $k\sim0.33$ \hMpc\ for all cosmologies. This indicates that the BAM-kernel effectively captures the scale-dependent features of the growth factor that appear in the power spectrum of the MG models with high accuracy.
\begin{figure*}
    \centering
    \includegraphics[width=\textwidth]{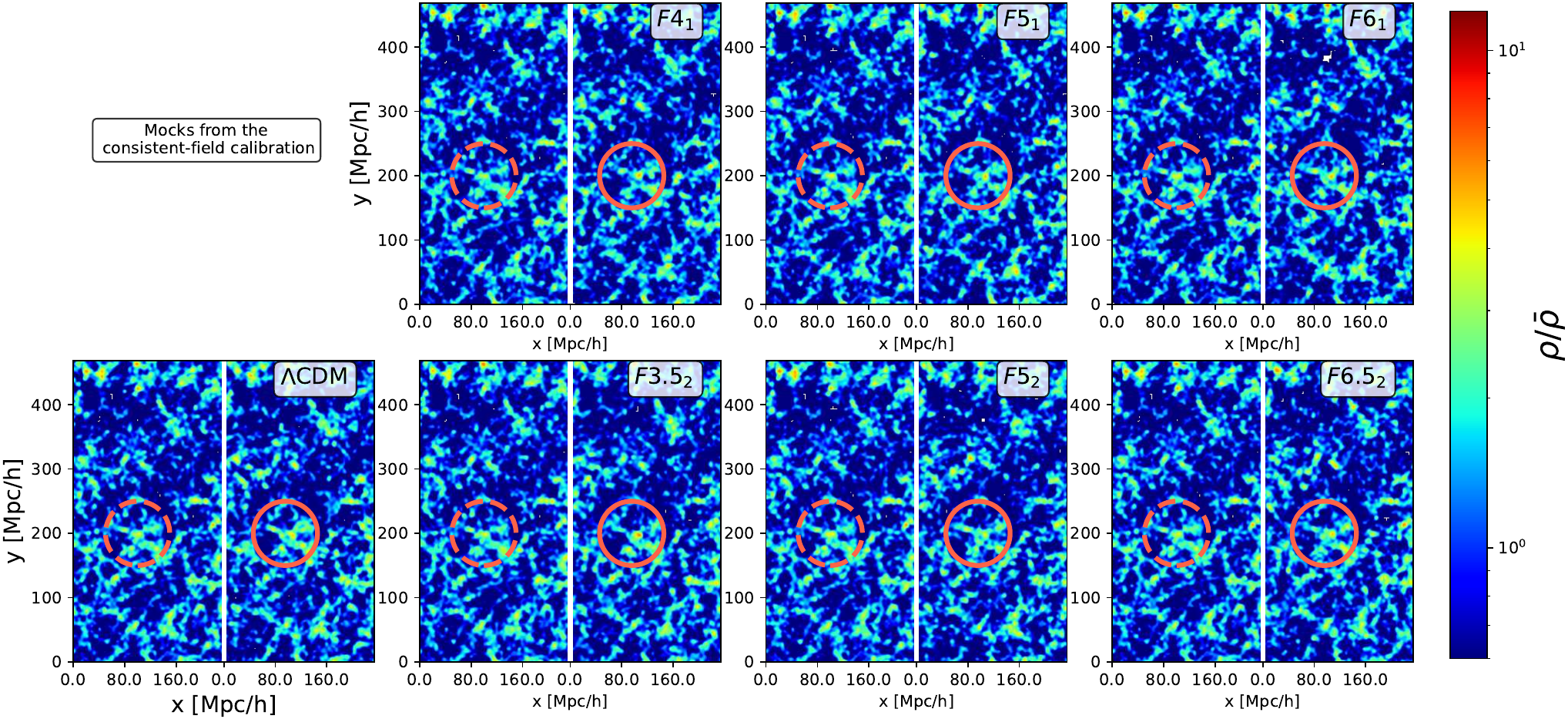}
    \includegraphics[width=\textwidth]{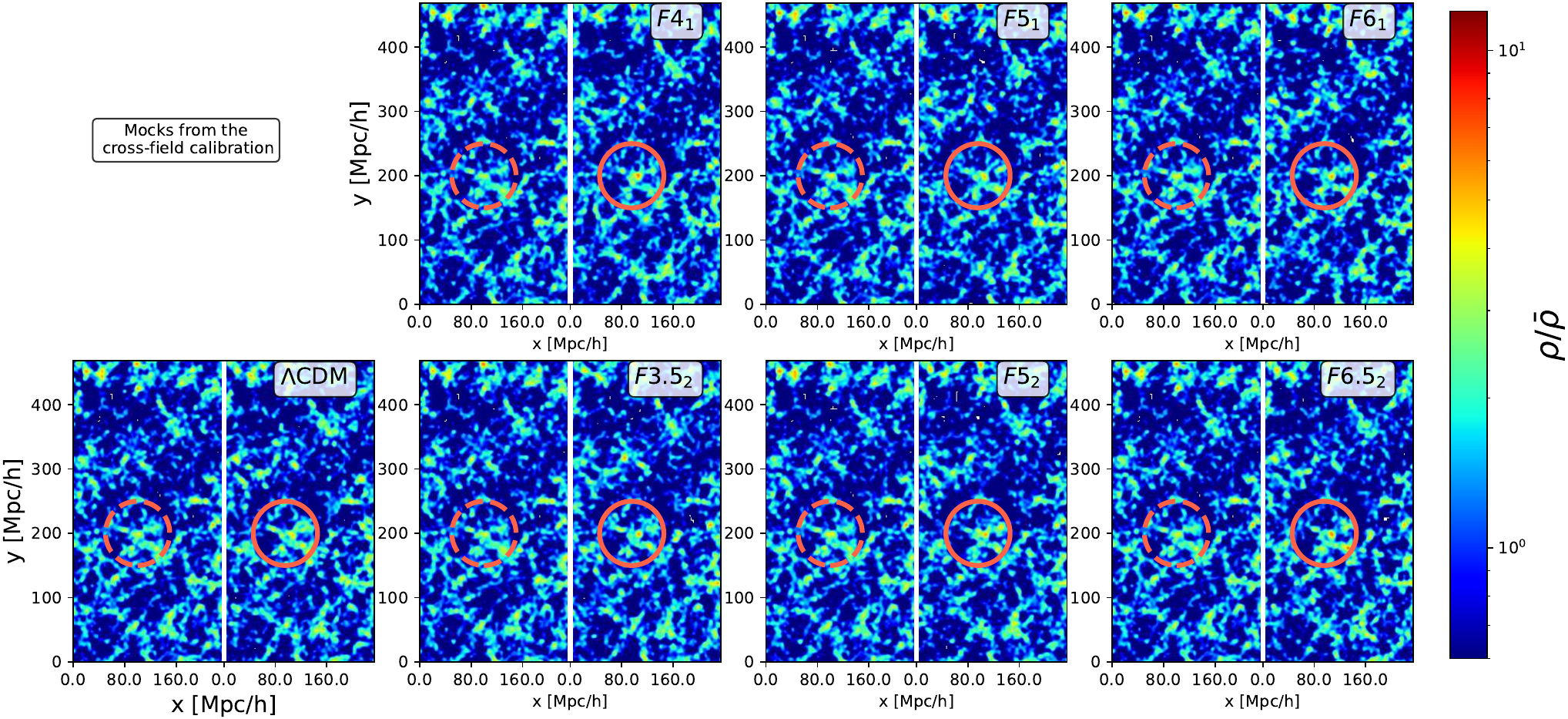}
    \caption{\small{Projected density field in slices of $240\times500\times98$ \Mpch\ from a volume of $1000^3$ \Mpch${}^3$ of the reference and mock catalogues for all cosmologies. Upper panels: mocks from the consistent-field calibration. Bottom panels: the mocks from the cross-field calibration. Each sub-panel is divided in two, the left side corresponds to the reference halos from the $N$-body simulation, while the right side of the sub-panel corresponds to their respective mock obtained with BAM. The models are as labelled and the color bar indicates the density $1 + \delta_{h}$. Note that the \lcdm\ model is a particular case in which the mock is the same in both upper and lower panels, as it serves as the reference model. The circles in each panel display a region of interest where visual differences in overdensity are visible to the naked eye (dashed circle for reference halos and solid circle for mocks).}}
    \label{fig:mockfields}
\end{figure*}

Figure \ref{fig:mockfields} displays the projected density field of the mock halo catalogues (right side of sub-panels) in contrast to their respective reference (left side of sub-panels). The slices are shown for all cosmologies and the two calibrations, labelled as mocks from the consistent-field calibration (upper panels) and mocks from the cross-field calibration (bottom panels). Although both fields (reference and mock) are notably similar, some subtle differences can be appreciated within the circles highlighted in each sub-panel. Nevertheless, in both sides of the panels the large-scale structure is clearly consistent.

A qualitative analysis of Fig. \ref{fig:mockfields} allows us to describe the region enclosed by the red circle in terms of its density variation compared to the reference catalogue. Overall, in the case of consistent-field calibration, $N^{\Vert}_h$, all models except $F5_2$ exhibit a slight excess in density. The mock of \lcdm\ shows no major deviations from its reference catalogue, which is consistent with previous analyses \citep[see e.g,][]{2019MNRAS.483L..58B, 2020MNRAS.493..586P, Balaguera_2020MNRAS, 2022MNRAS.512.2245K}. However, the mocks of the two extreme HS models, $F4_1$ and $F3.5_2$ which deviate significantly from GR gravity, display visible density peaks compared to their references. Conversely, the $F6_2$ model is almost indistinguishable, with results comparable to those of \lcdm. Notably, density peaks in the mocks are particularly visible in knots and sheets.
On the other hand, the mocks from the cross-field calibration, $N^{\times}_h$, exhibit a similar behaviour on large scales to those produced by the consistent calibration. However, in this case, when using the \lcdm\ DM density field to generate MG mocks, the density peaks appear higher, particularly in the circled region used for comparison, against to the consistent calibration. Specifically, the $F3.5_2$ model exhibits a slightly lower overdensity at the nodes compared to the previous calibration but remains distinguishable from the reference. Similarly, the $F4_1$ model displays a denser knot region than in the previous calibration, resulting in a significantly denser region than the reference catalogue. On the contrary, the $F5_1$ model shows comparable results in both calibrations, with no significant visual differences observed. The highlighted region in $F5_2$ has much more power compared to the previous calibration, making it clearly distinguishable from the reference. Furthermore, $F6_1$ displays slightly denser regions in that area compared to the consistent-field calibration, making it easier to distinguish from \lcdm. As in the previous model, $F6_2$ exhibits more power compared to the consistent calibration, resulting in clear differences from its reference catalogue, despite being the closest model to \lcdm\ in terms of clustering.

\begin{figure*}
    \centering
    \includegraphics[width=\textwidth]{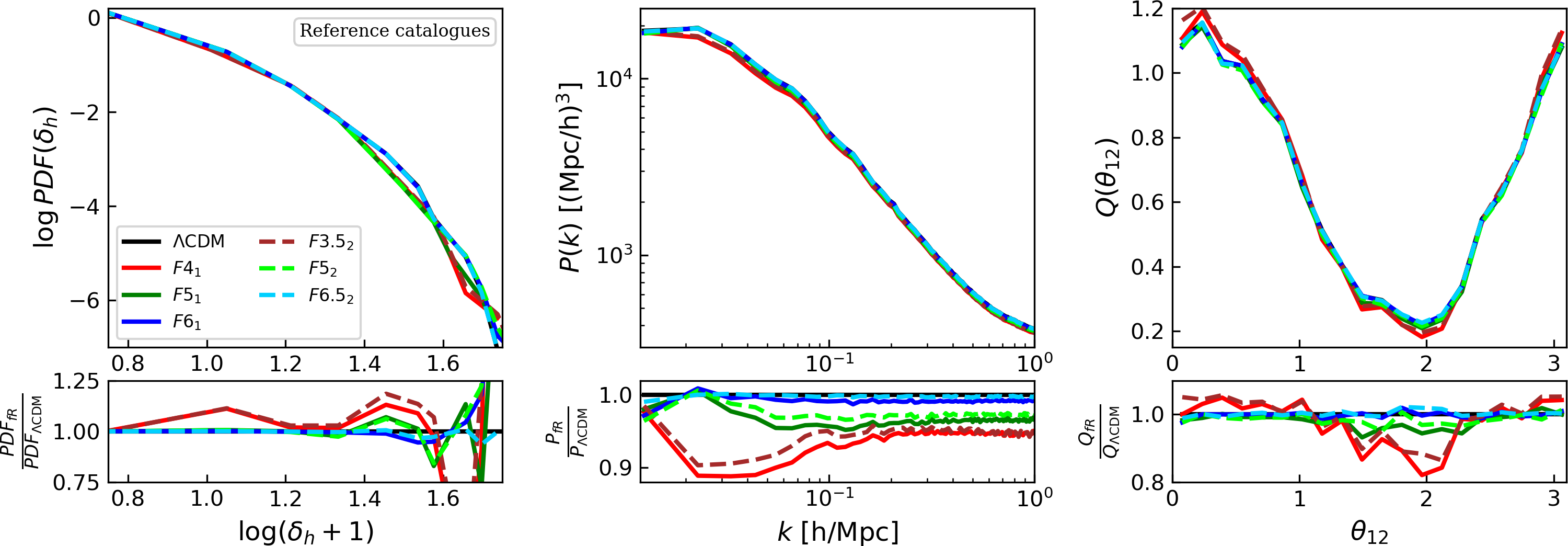}
    \includegraphics[width=\textwidth]{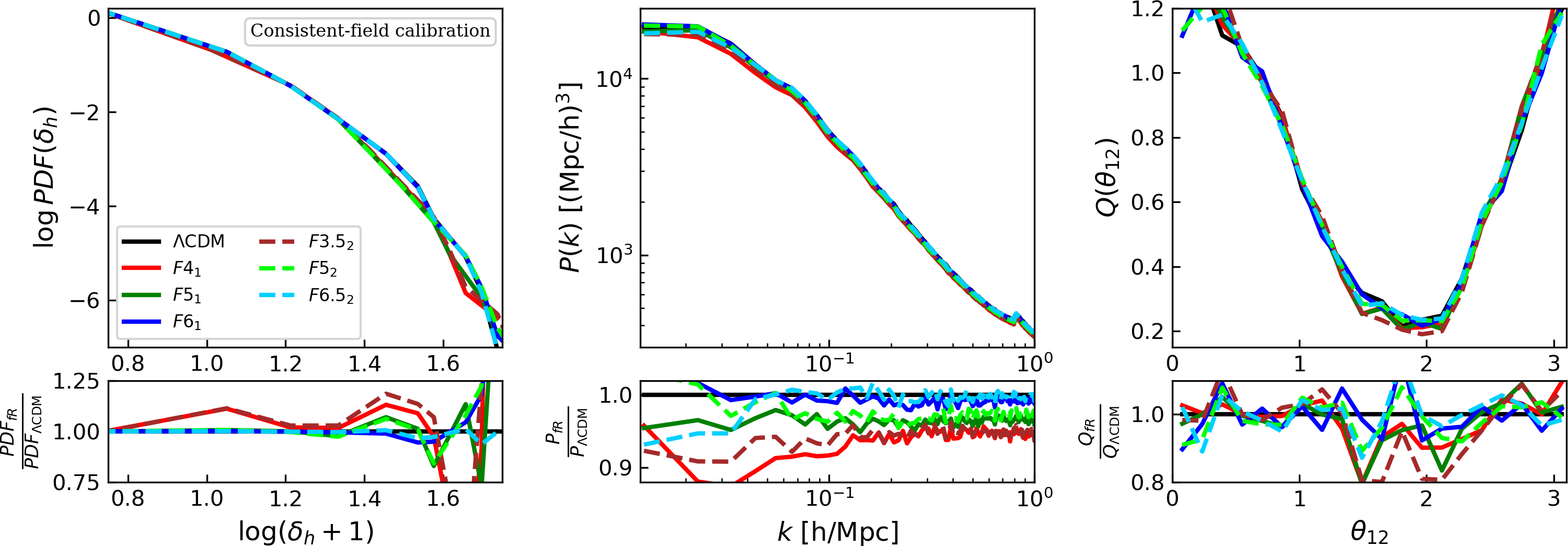}
    \includegraphics[width=\textwidth]{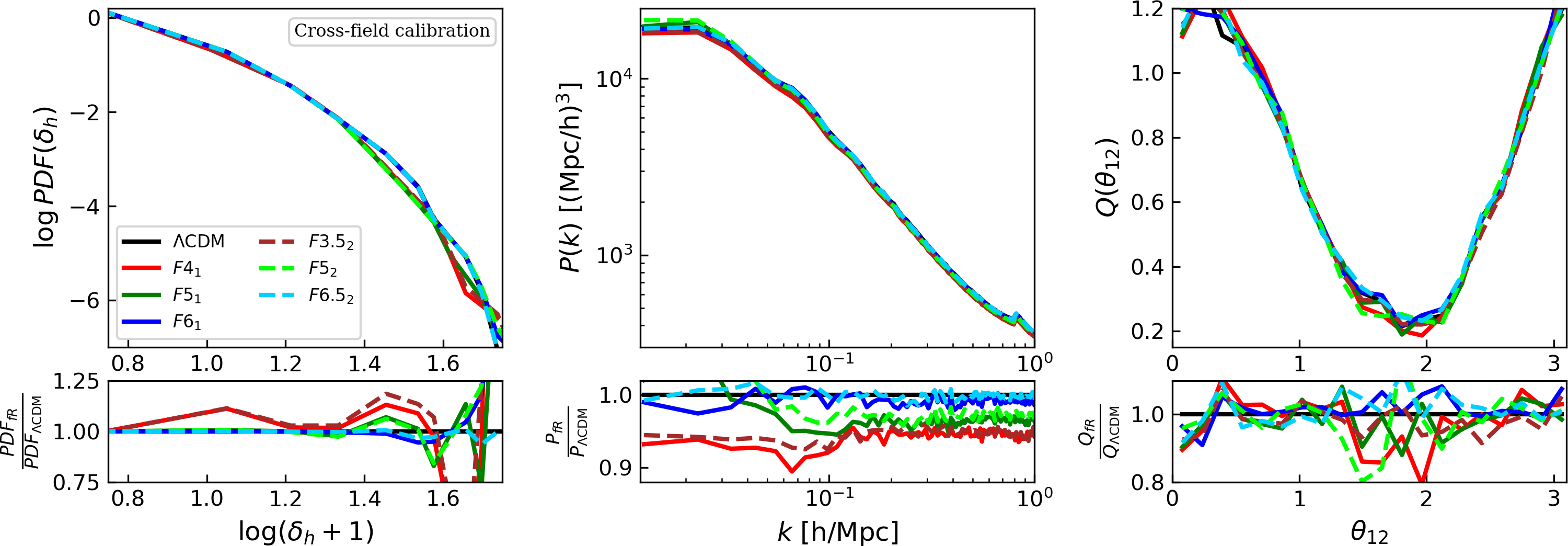}
    \caption{\small{Comparison of the probability density functions (PDF, left), power spectra ($P(k)$, middle) and reduced bispectra ($Q(k_1, k_2, \theta_{12})$, right) obtained from the calibration stage obtained with BAM for the different modified gravity models as labelled. Upper panels: reference halo catalogues. Middle row: mock calibration created with BAM from the consistent MG DM field of the each model. Lower row: mock calibration created with BAM from the DM field of the \lcdm\ model. Note that all the comparison are made between the MG model with respect to the $\Lambda$-CDM DM field.}}
    \label{fig:refcalibration_comp}
\end{figure*}
To evaluate the robustness of the calibrations with MG fields, we performed a comprehensive comparison of several summary statistics, including the PDF and the two- and three-point statistics of the halo distributions. We first compared the reference catalogues themselves, this is the MG models against the \lcdm\ model (\ie, $N^{\mathrm{ref}}_{h,\,\mathrm{MG}}$ vs $N^{\mathrm{ref}}_{h,\,\text{\lcdm}}$), aiming to describe the specific signatures of the HS models in the summary statistics that distinguish them from the standard model. We then perform a similar comparison for the consistent- and cross-field calibrations, which correspond to $N^{\Vert}_{h,\,\mathrm{MG}}$ vs $N^{\Vert}_{h,\,\text{\lcdm}}$, and $N^{\times}_{h,\,\mathrm{MG}}$ vs $N^{\times}_{h,\,\text{\lcdm}}$, respectively. This analysis holds significance as it provides insights into the degree of physical information encoded within the bias-relation $\mathcal{B}$ ($\tilde{\mathcal{B}}$) provided by \texttt{BAM}. It allows us to ascertain the effectiveness of these mapping relations in accurately reproducing nonlinear and non-local bias features, particularly in cosmologies characterised by scale-dependent growth factors. Fig. \ref{fig:refcalibration_comp} presents the outcomes of this analysis, with the mock comparison arranged from top to bottom as follows: reference catalogues only, consistent-field calibration, and cross-field calibration. The columns correspond, from left to right, to the PDF, power spectrum $P(k)$ and reduced bispectrum, $Q(\theta_{12}|k_1,k_2)$, in the particular configuration of $k_2=2k_1=0.2$ \hMpc. In this figure, all the ratios have been plotted with respect to the corresponding \lcdm\ halo catalogue.

The selected HS models are characterised by higher values of $|f_{R0}|$ that tend to manifest more pronounced deviations from \lcdm\ in terms of density fluctuations. Consequently, $F4_1$ exhibit more significant enhancements compared to $F5_1$ and $F6_1$ counterparts. This trend arises from the fact that greater $|f_{R0}|$ values correspond to more substantial modifications to gravity, which consequently exert a more pronounced influence on the cosmic structure formation. In the first row of Fig. \ref{fig:refcalibration_comp}, we observe the expected behaviour in the summary statistics of the reference MG halos compared to the \lcdm\ ones. Notably, the PDF of MG models featuring significant gravity modifications, such as $F3.5_{2}$ and $F4_{1}$, exhibit an excess of probability of up to 20\% of finding halos in cosmic environments whose overdensities are $\log(\delta_h+1)\approx1\;\text{and}\;1.5$. This trend extends to models with moderate deviations from GR, such as $F5_1$ and $F5_2$, albeit for denser environments with $\log(\delta_h+1)\approx 1.5$ to $1.6$. Conversely, models with weaker gravity modifications, such as the $F6_{1}$ and $F6.5_{2}$, closely resemble the PDF of \lcdm\ halos, with probability spreads below 2\% even at higher densities. The PDF variations also manifest in the two-point statistics, particularly in the suppression of power in the low $k$ modes. For instance, the suppression of power is better appreciated for $F4_1$ (exceeding 10\%), compared to $F3.5_2$ (around 10\%). Similarly, the $F5_1$ and $F5_2$ models show a more moderate power suppression, remaining below 5\% for $k<0.08$ \hMpc. As expected, the models with weak deviations from gravity, such as $F6_1$ and $F6_2$, display a consistent clustering across all scales when compared to \lcdm. Moreover, an effective bias between the MG and \lcdm\ models is observed for scales $k>0.1$ \hMpc, with this bias monotonically increasing as the intensity of gravity deviation grows. As for the three-point statistic, the reduced bispectrum, $Q(\theta_{12})$, is also sensitive to modifications of gravity, as can be seen by the changes in its characteristic U-shape. In fact, the deeper the inflection point, the more concave the bispectrum is, as seen for cosmologies with larger deviations from GR. In particular, when $\theta_{12}\approx\pi/2$, the $F3.5_2$ and $F4_1$ models exhibit deviations close to 20\% compared to \lcdm, while the $F5_1$ and $F5_2$ models show deviations below 5\%, and $F6_1$ and $F6_2$ closely resemble \lcdm\ for the entire range of $\theta_{12}$ values.

\begin{figure*}
    \centering
    \includegraphics[width=\textwidth]{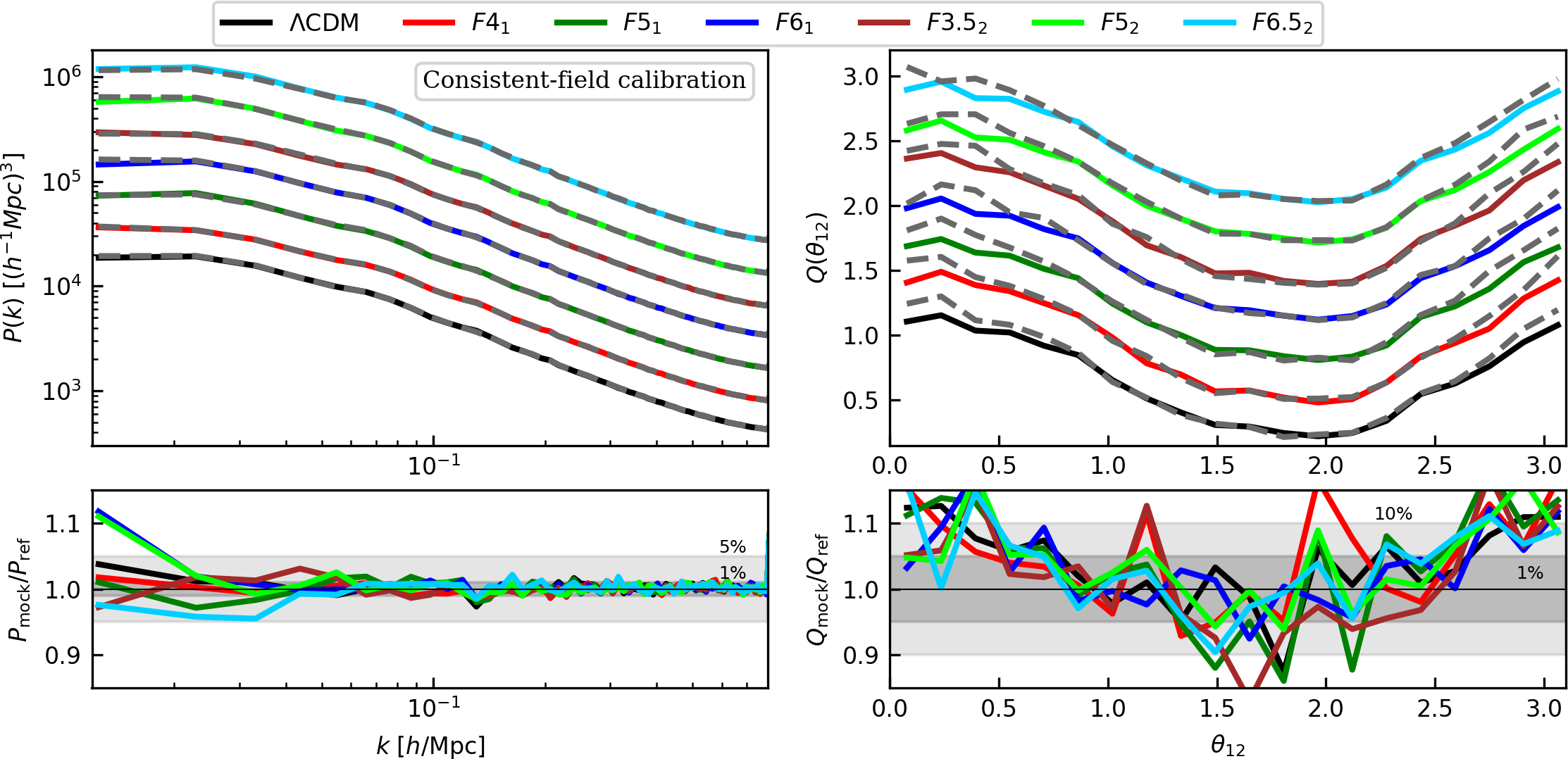}\bigskip\\
    \includegraphics[width=\textwidth]{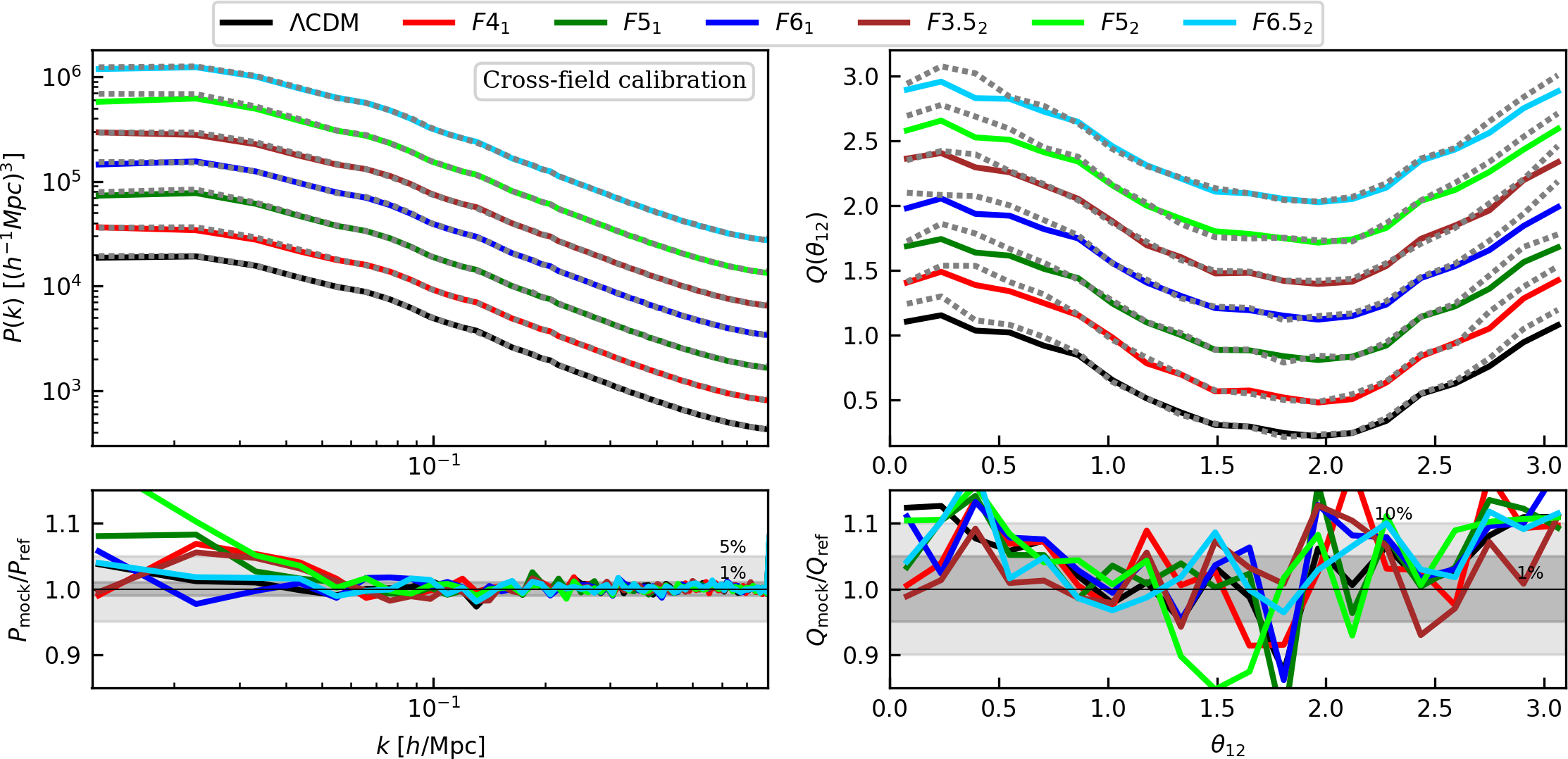}
    \caption{\small{Power spectrum and reduced bispectrum of the different MG mocks compared to their respective reference catalogues for the studied calibrations as labelled. Upper panels: consistent-field calibration, \ie, $N^{\Vert}_{h,\,\mathrm{MG}}$ vs $N^\mathrm{ref}_{h,\,\mathrm{MG}}$. Lower panels: cross-field calibration, \ie, $N^{\times}_{h,\,\mathrm{MG}}$ vs $N^\mathrm{ref}_{h,\,\mathrm{MG}}$.}}
    \label{fig:summarystatistics2}
\end{figure*}

The second and third rows of Fig. \ref{fig:refcalibration_comp} show the summary statistics obtained from the consistent-field calibration, $\delta_{\mathrm{DM}}^{\mathrm{MG}}$\SymbolBiasEffective$N_h^{\mathrm{MG}}$, and cross-field calibration, $\delta_{\mathrm{DM}}^{\text{\lcdm}}$\SymbolBiasEffective$N_h^{\mathrm{MG}}$, respectively. Analogous to the results of the references catalogues (see first row of Fig. \ref{fig:refcalibration_comp}), we focus on the mock quality regarding the signatures captured by the \texttt{BAM} bias relation with respect to \lcdm\ model rather than its reference catalogue, \ie, $N^{\Vert}_{h,\,\mathrm{MG}}$ vs $N^{\Vert}_{h,\,\text{\lcdm}}$, and $N^{\times}_{h,\,\mathrm{MG}}$ vs $N^{\times}_{h,\,\text{\lcdm}}$.

First of all, we observe that the PDF of all MG models remain identical in both calibrations, mirroring their respective reference catalogues. This consistency is an inherent feature of \texttt{BAM} method. Moreover, the mocks of both calibration approaches reproduce remarkable well the power spectrum of the references catalogues beyond scales $k\geqslant0.1$ \hMpc, as evidenced by the ratios of the power spectra. That is, the non-linear information encoded in $N_{h,\,\mathrm{MG}}$ is effectively captured by the bias relationship and \texttt{BAM}-kernel of both calibrations. Below this scale ($k<1$ \hMpc), however, we find deviations in the power spectrum of MG models, primarily attributed to cosmic variance stemming from the finite number of independent modes sampled within the grid volume. In this respect, the $F3.5_2$ and $F4_1$ mocks of the consistent-field calibration faithfully reflect the behaviour of the references on the largest scales, while the remaining  models become entangled within the cosmic variance, showing a clustering suppression ranging between approximately $4\%$ and $8\%$ in all cases, even in those models that closest resembles the \lcdm\ DM field, such as $F6_1$ and $F6.5_2$. On the other hand, the power spectrum of the cross-field mocks aligns well with that of the references on large scales. Both families of models -- $F5_1$, $F5.5_2$ and $F6_1$, $F6.5_2$ -- show a closer resemblance to the clustering of the \lcdm\ mock, while mocks from $F3.5_2$ and $F4_1$ exhibit noticeable deviations of up to $8\%$. However, it is worth noting that for the latter models, the \texttt{BAM} calibration tends to keep a consistent clustering suppression, unlike the characteristic patterns observed in the power spectrum of the reference catalogues (see first row of Fig.~\ref{fig:refcalibration_comp}). This result can be attributed to the fact that the \texttt{BAM}-kernel primarily corrects clustering on non-linear scales rather than linear scales, where the power spectrum is expected to be easily mapped. At the same time, on these scales, the effects of MG become more significant in the power spectrum because, unlike on small scales where the chameleon mechanism is very efficient, here it tends to be weaker. Given that cosmic variance affects all models equally at the largest scales, we consistently observe the same hierarchy in the relative differences of the power spectrum compared to the \lcdm\ mocks, \ie, the features of MG are preserved with comparable accuracy to those of the original reference catalogues.

Last column of Fig.~\ref{fig:refcalibration_comp} shows the differences in the reduced bispectrum for both calibrations. Quantifying the discrepancies in the bispectrum proves challenges owing to the subtle deviations from \lcdm\ observed in the reference catalogues. When translated to the calibrations, most of the deviations manifest in the depth of the U-shape around $\theta_{12}=\pi/2$. Surprisingly, in both calibrations, the U-shape of models $F5_1$ and $F6_1$ is enhanced by an oscillating behaviour around the fiducial value set by \lcdm. $F3.5_2$ and $F4_1$ mocks remain distinguishable in both calibrations, although they do not follow the same trend as marked by the reference catalogues. The deviations in the bispectrum wings are within $5\%$, for both, consistent and cross-calibrations, being in agreement with previous studies \citep[see \eg,][]{2015MNRAS.451..539G,2020MNRAS.493..586P,2022MNRAS.512.2245K}. The reduced bispectrum indicates that while some of the MG features are not entirely recovered in the U-shape of $F5_1$ and $F6_1$ models, the rest of MG models remain distinguishable from \lcdm. Possible factors explaining this include systematic errors in the \texttt{BAM} calibrations, resolution effects due to the employed grid-size, and the intrinsic nature of the chameleon mechanism. On this latter aspect, \citet{2011JCAP...11..019G} demonstrated the sensitivity of the reduced bispectrum to the chameleon mechanism, revealing substantial deviations between MG models and \lcdm, reaching up to 10-15\%.

To move forward with our analysis and assess the mock quality and efficacy of the bias mapping of \texttt{BAM} in MG cosmologies, we compare the MG mocks to their corresponding reference catalogues, denoted as $N^{\Vert}_{h,\,\mathrm{MG}}$ vs $N^\mathrm{ref}_{h,\,\mathrm{MG}}$, and, $N^{\times}_{h,\,\mathrm{MG}}$ vs $N^\mathrm{ref}_{h,\,\mathrm{MG}}$. Previous studies \citep[][]{2019MNRAS.483L..58B, 2020MNRAS.493..586P, Balaguera_2020MNRAS, 2022MNRAS.512.2245K} demonstrated that \texttt{BAM} achieves accuracy within 1\% in the power spectrum for \lcdm\ simulations. Therefore, our comparison is particularly insightful as it allows to evaluate \texttt{BAM}'s accuracy at reproducing the halo distribution in a new scenario, namely, using MG cosmologies. In that regard, Fig.~\ref{fig:summarystatistics2} shows the power spectrum and bispectrum of the MG mocks for both the consistent-field calibration (upper panels) and cross-field calibration (lower panels). Both the power spectra and bispectra curves have been shifted for clarity. Note that the \lcdm\ model is special, as its statistics remain the same in both calibrations. The two calibrations (consistent- and cross-calibration) show excellent agreement with the reference power spectrum, with residuals between the mock power spectra and the reference one  within 1\% up to $k\sim 0.8$ \Mpch. These precision underscore the importance of considering both local and non-local properties of the reference catalogue \citep{2022MNRAS.512.2245K, 2024A&A...685A..61B}.  

The precision achieved in the power spectra ratios of the consistent-field calibration is expected, given that the halo distribution inherits the non-linear and non-local properties of the respective MG DM density field. Therefore, it is not surprising that the range of accuracy is broader in this calibration, as depicted in Fig.~\ref{fig:summarystatistics2}. Regarding the cross-field calibration, we observe the anticipated deviations in the power spectrum at large scales, arising from the lack of power of the MG models compared to \lcdm, as previously discussed in Fig.~\ref{fig:refcalibration_comp}. The discrepancies in the power spectrum of the MG mocks at scales below $k<0.07$ \hMpc correspond to specific MG features of the HS models. In particular, we notice that the suppression of power in the halo distribution of the $F3.5_2$ and $F4_1$ mocks manifests in the lower panels of Fig.~\ref{fig:summarystatistics2} as a bump in power at the same scales. Similarly, the mocks for $F5_1$ and $F5.5_2$ exhibit an excess in power of approximately 10\% at scales $k\approx0.02$ \hMpc. The residuals between the $P(k)$ of $F6_1$ and $F6.5_2$ and their references are within 5\%, even at the largest scales, closely resembling the behavior of the \lcdm\ halos, \emph{which is a good indicator that even when calibrations are performed using crossed DM fields, it is feasible to derive statistics of the MG halo distribution}. By analysing the three-point statistics of the consistent-field calibration, we observe a U-shaped pattern in the residuals of the bispectra of the mocks. This indicates a better fit at the inflection point, while improvement is desirable on the wings. However, this shape is much reduced in the cross-field calibration, which is due to a minor increase in the agreement of the inflection point, while the fit of the wings stays consistent between calibrations. Our findings show that the mock of $F3.5_2$ in the cross-field calibration improves the inflection point somewhat compared to the consistent-field calibration, while residuals remain within 10\% in both calibrations. Likewise, the $F4_1$ mocks exhibit minimal deviation in the bispectra residuals, with both calibrations showing consistency within 10\% across most of the $\theta_{12}$ range. Mocks of the $F5_1$ model do show a slight improvement at $\theta_{12}\approx\pi/2$, but this improvement holds true for both calibrations. As compared to the cross-field calibration, the mocks of $F5_2$ demonstrate improved agreement in consistent-field calibration across the entire $\theta_{12}$ range, reducing deviations by over 10\%. Last but not least, the mocks of $F6_1$ and $F6.5_2$ cosmologies show good agreement with their reference catalogues in both calibrations, with only one point deviating at $\theta_{12}\approx1.8$.

The analysis of the reduced bispectrum in HS models uncovers deviations of up to 10-15\% compared to \lcdm, highlighting its potential to discern between MG models. In particular, MG cast a weaker influence on the bispectrum than on the power spectrum, suggesting its utility in breaking galaxy-bias degeneracies. The consistent differences observed in both the power spectrum and the bispectrum emphasise the critical importance of accurate measurements for model constraints. Actually, the bispectrum provides a valuable diagnostic tool for delineating MG effects and warrants further exploration through various parameter combinations to maximise the signal-to-noise ratio. It is worth noting that within our analysis, the bispectrum is not a calibrated quantity in the \texttt{BAM} approach, which underscores that the observed agreement, as shown in Fig. \ref{fig:summarystatistics2}, is naturally inherited from the local and non-local properties of the density fields used in the multi-dimensional halo bias described in \S\ref{sec:app}.

\section{Summary and discussion}\label{sec:summaryconclusions} 
This study focuses on efficiently generating modified gravity (MG) catalogues from mapping either standard or MG-dark matter density fields using the bias assignment method (\texttt{BAM}). Our results assess the flexibility of \texttt{BAM} in effectively modelling the effects of MG using a benchmark training data catalogue, by incorporating non-local and non-linear information in the description of halo bias. The analysis was conducted across six distinct cosmologies based on the Hu-Sawicki (HS)  \citep{HuSawicki_2007} parametrisation of the $f(R)$ gravity model. These cosmologies encompass various levels of deviation from the \lcdm\ model, including those capable of mimicking the clustering of \lcdm, as well as those with significant deviations that have already been ruled out by observations solar system constraints. One important component of our study is the exploration of the growth factor and effective gravitational constant for different configurations of the power-law index, $n$, and the magnitude of the scalar field $|f_{R0}|$ of the MG models. Both the scale-dependent growth factor and variations in gravity are crucial for understanding bias relations in MG models. Therefore, our results focus on the redshift $z=0.5$ to align with recent galaxy survey data and span a wide range of $k$-modes in Fourier space, providing comprehensive coverage across both large and small scales. In particular, we consider the distribution of dark matter halos generated in different MG models generated with the \texttt{COLA} , and used \texttt{BAM} to obtain non-parametric bias relations for each MG model. We employ two distinct calibration experiments, namely  consistent-field and cross-field calibrations. In the consistent-field calibration, we use the DM of MG models to mimic the corresponding MG catalogues, ensuring self-consistency between DM and number counts of biased tracers. This approach is expected to effectively capture most of the MG effects in the bias relation. Conversely, the cross-field calibration offers the opportunity to rapidly generate mock MG catalogues by using a mapping relationship based on the DM density field of a \lcdm\ simulation rather than running MG simulations, which are typically more demanding.

We employ the two- and three-point statistics to assess the effectiveness of the mappings relations obtained with our non-parametric approach. It should be stressed that during the calibration process, the one-point statistic namely the PDF, is adjusted to match the desired number counts, in our case those of the MG models. In the mean time, the power-spectrum is computed for each iteration, during which the DM is convolved with the \texttt{BAM}-kernel until it reaches sufficient accuracy to reproduce the target power spectrum of the MG halos. It is noteworthy that the three-point statistic is not directly involved in the calibration process; therefore, the results obtained reflects the amount of information  in the non-linear and non-local information of the density field \citep[see e.g.][]{2019MNRAS.483L..58B,Balaguera_2020MNRAS,2022MNRAS.512.2245K,2024arXiv240500635B}. Our results demonstrate that the MG halos obtained from both calibrations excel in summary statistics, achieving a 1\% accuracy in the power spectrum across a wide range of $k$-modes, with only minimal differences well below 10\% at low modes, particularly below $k<0.07$ \hMpc. Meanwhile, the bispectrum remains consistent with the reference catalogues within 10\% for all values of the $\theta_{12}$ angle in the typical configuration $k_2=2k_1=0.2$ \hMpc. However, model-specific discrepancies for each MG cosmology become apparent when examining bispectrum values below 5\%, revealing more intricate behaviour that, in general, depends on the complex growth of structure in these models, as detailed in \S \ref{subsec:performance_statistics}.

The analysis of \texttt{BAM} calibrations revealed that the consistent-field demonstrated a good accuracy over broader ranges, showcasing the non-linear and non-local properties inherited from the MG DM density fields. In contrast, the cross-field calibration display the predicted deviations on large-scales due to differences in power between MG models and the standard \lcdm\ model. Specific features of the HS parametrisations led to deviations in the power spectrum (with some models showing suppression thereof), especially at large scales. Moreover, deviations of up to 10-15\% in the reduced bispectrum compared to \lcdm\ highlighted the potential of this statistic to distinguish between MG models and its potential in breaking galaxy-bias degeneracies. The consistent-field calibration showed a U-shaped pattern in bispectrum residuals, indicating that a better fit at the inflection point is desirable, but there is room for improvement on the wings, while the cross-field calibration exhibited flat ratios with consistent wing fit between calibrations.

The precision achieved when using MG reference catalogues demonstrates the competitive capability of non-parametric architectures, such as \texttt{BAM}, in generating mocks from DM fields, whether it is \lcdm\ or MG. In fact, learning complex bias relations is essential for generating mock catalogues that faithfully reproduce the halo number counts and statistical features of cosmologies beyond \lcdm, facilitating further cosmological analysis. As mentioned before, a key factor that allows us to achieve unprecedented accuracy is the inclusion of non-linear and non-local information from the reference training data set via the cosmic web classification. In the MG context, this is even more relevant due to the nature of the chameleon mechanism, which operates in high-density environments by suppressing modifications to gravity while allowing significant modifications on cosmological scales. This highlights the fact that the HS scalar field leads to amplified growth of structures on large scales, resulting in higher-density peaks compared to the \lcdm\ model. In contrast, on smaller scales, these modifications tend to decrease structure formation, leading to a smoother density field. Indeed, in dense environments like clusters, the chameleon mechanism operates efficiently, leading to suppressed gravity modifications, while in under-dense regions such as voids, the chameleon mechanism is ineffective, resulting in significant modifications to gravity. \citet{2012MNRAS.421.3481L} demonstrates that this environmental dependence serves as strong evidence for gravity modification in models featuring the chameleon mechanism. Thus, by employing a non-linear and non-local bias description, our results confirm that the effects of MG can be effectively captured when environmental information of the DM density field is provided.

Overall, this approach enhances the representation of the matter distribution in alternative cosmological scenarios, with diverse structure formation and gravitational effects across different scales. The results indicate that summary statistics, including power spectra and reduced bispectrum, align with the reference catalogues for various MG models. This suggests that \texttt{BAM} accurately reproduces the clustering patterns and deviations caused by MG, providing valuable insights for future cosmological analyses. Looking ahead, we plan to assess the impact of different MG models in the scaling relations of the properties of dark matter halos, their correlations with the large-scale effective primary and secondary bias \citep[][]{2024A&A...685A..61B}. We will also apply the parametric mapping technique \citep[][]{2024arXiv240319337C} to each of the different regions of a novel hierarchical cosmic web classification. These developments and insights on the clustering in non standard cosmologies promises to further refine the MG mock catalogue generation process and will be useful in the context of testing gravity in surveys such as EUCLID, DESI, and LSST.  

\begin{acknowledgements}
We would like to thank Ismael Ferrero, Farbod Hassani, David Mota and Hans Winther for discussions in the early stage of this project. The authors acknowledge support from the Canarian government under the project ``Mapping the Universe with modified gravity'' PROID2021010126, and the Spanish Ministry of Economy and Competitiveness (MINECO) for financing the \texttt{Big Data of the Cosmic Web} project: PID2020-120612GB-I00/AEI/10.13039/501100011033, and the IAC for continuous support to the \texttt{Cosmology with LSS probes} project. JEGF is supported by Spanish Ministry of Universities, through a Mar\'ia Zambrano grant (program 2021-2023) at Universidad de La Laguna with reference UP2021-022, funded within the European Union-Next Generation EU. ABA acknowledges the Spanish Ministry of Economy and Competitiveness (MINECO) under the Severo Ochoa program SEV-2015-0548 grants. We also thank the personnel of the Servicios Inform\'aticos Comunes (SIC) of the IAC.
\end{acknowledgements}

\bibliographystyle{aa}
\bibliography{bibliography}
\end{document}

%% file: newcomands.tex
\newcommand{\lcdm}{$\mathrm{\Lambda CDM}$}
\newcommand{\Mpch}{$h^{-1}\,\mbox{Mpc}$}

\newcommand{\hMpc}{$h\,\mbox{Mpc}^{-1}$}
\newcommand{\Msunh}{$h^{-1}\,\mbox{M}_\odot$}

\newcommand{\ie}{\textit{i}.\textit{e}.}
\newcommand{\eg}{\textit{e}.\textit{g}.}

\newcommand{\SymbolBiasEffective}{%
  \begin{tikzpicture}[baseline={(B.base)}]
    \node[inner sep=0pt, outer sep=0pt] (B) {{\scriptsize$\tilde{\mathcal{B}}$}};
    \draw (B.north) ++(0.15em,0.25em) -- ++(-0.15em,-0.15em);
    \draw (B.north) ++(0.15em,0.25em) -- ++(-0.15em,0.15em);
    \draw (B.south) ++(-0.1em,-0.25em) -- ++(0.15em,0.15em);
    \draw (B.south) ++(-0.1em,-0.25em) -- ++(0.15em,-0.15em);
    \draw (B.west) ++(-0.35em,0) -- ++(-0.6em,0);
    \draw (B.east) ++(0.35em,0) -- ++(0.6em,0);
    \draw (B.center) circle (0.6em);
  \end{tikzpicture}%
}